\documentclass[a4paper,12pt]{article}
\usepackage{amsmath,feynmf,amsfonts,amssymb}
\usepackage{booktabs}
\usepackage{graphicx}
\usepackage{lscape,epsf}

\newcommand{\ME}[3]{\langle #1 | #2 | #3 \rangle}

\newcommand{\tr}{{\rm tr \ }}
\def\ds{\displaystyle}
\def\bea{\begin{array}{c}}
\def\ea{\end{array}}
\def\be{\begin{equation}\bea\ds}
\def\ee{\ea\end{equation}}

\newcommand\Trule{\rule{0pt}{2.6ex}}
\newcommand\Brule{\rule[-1.2ex]{0pt}{0pt}}

\def\Fc{{\cal{F}}}

\def\Leff{${\cal L}_{{\rm eff}}$}

\begin{document}\setlength{\unitlength}{1mm}

\begin{titlepage}

\begin{flushright}
{RUNHETC-2008-18}\\
{TAUP-2890-08}\\
{ITEP-TH-45/08}\\
\end{flushright}

\begin{center}
{\large {\bf A Pure-Glue Hidden Valley \\
\vskip 0.3cm
I. States and Decays}} \\
\vskip0.5cm {J.~E.~Juknevich$^{a}$, D.~Melnikov$^{b,c}$ and M.~J.~Strassler$^{a}$}
\vskip1cm
{\itshape $^{a}$Department of Physics and Astronomy,\\
Rutgers, 136 Frelinghuysen Rd, Piscataway, NJ 08854, USA\\
\vspace{0.5cm}
$^{b}$Raymond and Beverly Sackler School of Physics and Astronomy,\\
Tel Aviv University, Ramat Aviv 69978, Israel\\
\vspace{0.5cm}
$^{c}$Institute for Theoretical and Experimental Physics,\\
B.~Cheremushkinskaya 25, 117259 Moscow, Russia\\
}
\end{center}

\vspace{0.5cm}
\begin{abstract}
It is possible that the standard model is coupled, through new massive
charged or colored particles, to a hidden
sector whose low energy dynamics is controlled by a pure Yang-Mills
theory, with no light matter.  Such a sector would have numerous
metastable ``hidden glueballs'' built from the hidden gluons.
These states would decay  to particles
of the standard model.  We consider the phenomenology of this
scenario, and find formulas for the lifetimes and branching ratios of
the most important of these states.  The dominant decays are to two
standard model gauge bosons, or by radiative decays with photon
emission, leading to jet- and photon-rich signals.
\end{abstract}

\end{titlepage}

\tableofcontents


\section{Introduction}
With the Large Hadron Collider (LHC), particle physics enters a new
era of potential discovery, one which may provide insights into the
many puzzles of the standard model (SM).  Given the immense challenges
of hadron collider physics, and the degree to which the future of
particle physics rests on the LHC, it is important to ensure that the
LHC community is fully prepared for whatever might appear in the data.
This requires consideration of a wide variety of models and signatures
in advance of the experimental program.

There are a number of reasonable and motivated solutions to the
hierarchy problem, the most favored being supersymmetry (SUSY), with
others including the little Higgs, warped extra dimensions and
technicolor.  Most experimental studies of these ideas have focused on
``minimal'' models.  But unlike supersymmetry or the little Higgs, {\it minimality} does not by itself solve any particle
physics problem.  Instead, it is motivated by theorists' philosophical
attachment to beauty and elegance.
Given our experience with the standard model, whose third generation and
neutral currents were not viewed as well-motivated before they were
discovered, it is prudent that we examine non-minimal models as well,
to ensure that their signatures would not be missed at the LHC.  This is especially important because a small addition to a theory can lead to a large change in its LHC phenomenology.

One likely possibility for new non-minimal physics involves the presence
of a hidden sector with TeV-scale couplings to the standard model.  A
large fraction of these models fall within the ``hidden valley
scenario'' \cite{SZ,HV2,HV3,HVun,HVWis,hvstudy1}.
In the hidden valley scenario, a new hidden sector (the ``hidden
valley sector'', or ``v-sector'' for short) is coupled to the SM in
some way at or near the TeV scale, and the v-sector's dynamics
generates a mass gap.  Such a mass gap,\footnote{or more generally a
mass ``ledge'' where one or more new particles in the hidden sector,
unable to decay within its own sector, is forced to decay via its weak
coupling to the SM sector} independent of the dynamics leading to that
gap, ensures that there are particles that are stable or metastable
within the v-sector.  These can only decay, if at all, via their very weak
interactions with the SM.  Processes that access the hidden valley are
often quite unusual compared to those in minimal supersymmetric or
other well-studied models.  Production of v-sector particles commonly
leads to final states with a high multiplicity of SM particles.  Also,
a hidden valley often leads to particles that decay with macroscopic
decay distances.  The resulting phenomenological signatures can be
difficult, or at least subtle, for detection at the Tevatron or LHC;
see for example  \cite{SZ,HV2,hvstudy1}.

Hidden valleys have arisen in bottom-up models such as the twin-Higgs
and folded-supersymmetry models \cite{TwinHiggs,FoldedSUSY} that
attempt to address the hierarchy problem, and in a recent attempt to
explain the various anomalies in dark-matter searches
\cite{hvdarkmatter} which requires a dark sector with a new force and
a 1 GeV mass scale.  They are also motivated by top-down model
building: hidden sectors that are candidate hidden valleys arise in
many string theory models, see for example \cite{string hv}.  
In recent years string theorists have found many models
that apparently have the minimal supersymmetric standard model
as the chiral matter of the theory, but
which typically have extra vector-like matter and extra gauge groups.
The non-minimal particles and forces which arise in these various
models may very well be visible at the LHC \cite{SZ}.

In this paper we consider a hidden valley that at low energy is a
pure-Yang-Mills theory, a theory that has its own gluons
(``v-gluons'') and their bound states (``v-glueballs'').  This
scenario easily arises in models; for example, in
many supersymmetric v-sectors, supersymmetry breaking and associated
scalar expectation values may lead to
large masses for all matter fields.

\begin{figure}[htb]
\begin{center}

\epsfxsize=3.5in \epsffile{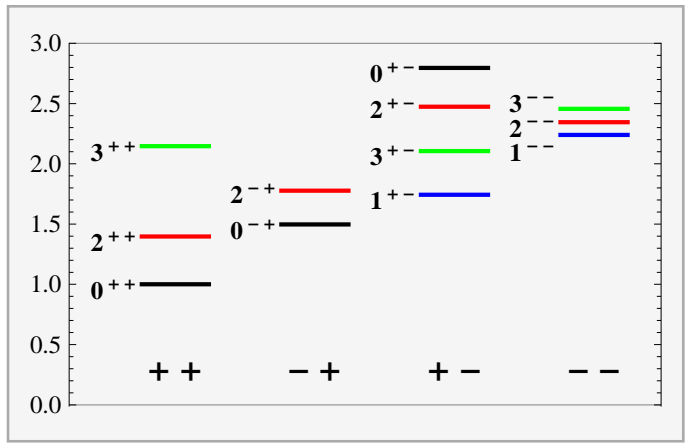}

\end{center}
\vspace{-0.6cm}
\caption{\small Spectrum of stable glueballs in pure glue $SU(3)$ theory \cite{Morningstar}.}
\label{fig spectrum}
\end{figure}

 The spectrum of stable
bound states in a pure Yang-Mills theory is known, to a degree, from
lattice simulations \cite{Morningstar}.  The spectrum of such states
for an $SU(3)$ gauge group is shown in figure \ref{fig spectrum}.  The
spectrum includes many glueballs of mass of order the confinement
scale $\Lambda_v$ (actually somewhat larger), and various $J^{PC}$
quantum numbers. All of the states shown are stable against decay to
the other states, due to kinematics and/or conserved quantum numbers.

In this paper we
will further specialize to the case where the coupling between the
SM sector and the v-sector occurs through a
multiplet of massive particles (which we will call $X$) charged under
both SM-sector and v-sector gauge groups.\footnote{Recently such
states, considered long ago \cite{Okun,quinn}, have been termed ``quirks'';
some of their very interesting dynamics, outside the regime we
consider here, have been studied in \cite{KLN}.}  A loop of
$X$ particles induces dimension-$D$ operators of the form
\begin{equation}
\label{loop terms}
 \frac{1}{M^{D-4}} {\cal O}_s^{(D-d)}{\cal O}_v^{(d)}
\end{equation}
 where $M$ is the mass of the heavy particle in the loop.  Here we
have split the dimension-$D$ operator ${\cal O}^{(D)}$ into a
Standard-Model part ${\cal O}_s^{(D-d)}$ of dimension $D-d$ and a
hidden-valley part ${\cal O}_v^{(d)}$ of dimension $d$.  All
v-glueball states can decay through these operators.

By simple dimensional analysis, these operators yield partial decay
widths of order $\Lambda_v^{2D-7} / M^{2D-8}$.  We will see that the
v-glueball decays are dominated by $D=8$ operators.  The next
operators have $D=10$, and their effects are typically suppressed by
$\sim (\Lambda_v/M)^4$.  The $D=8$ operators induce lifetimes for the
v-glueballs of order $M^8/\Lambda^9$, which can range anywhere from
$10^{-20}$ seconds to much longer than a second, depending on the
parameters.  Implicitly our focus is on the case where the lifetimes
are short enough that at least a few decays can be observed in an LHC
detector.  This typically requires lifetimes shorter than a
micro-second, if the production cross-section is
substantial.\footnote{To avoid any confusion, we emphasize again that
these v-glueballs have extremely weak interactions with the standard
model, and do not interact with the detector (in contrast to
R-hadrons, which are made from QCD-colored constituents and
have nuclear-strength interactions.)  They can only
be detected directly through their decay to standard model particles.}
However, our formulas will be valid outside this regime as well.

We will need to construct the $D=8$ effective action coupling the two
sectors.  Then we will use it to compute formulas for the partial widths
of various decay modes of the v-glueballs, concentrating on the lighter
v-glueball states, which we expect to be produced most frequently.

Application of our formulas, particularly as relevant for the LHC, will
be carried out in a second paper \cite{vglue2}.  To put the present
paper in context, we now briefly review the results to be presented there.
Although there are some irreducible uncertainties due to unknown
glueball transition matrix elements and decay constants, we find that
the various v-glueball states have lifetimes that probably span 3 or 4
orders of magnitude.  We also find that the dominant v-glueball decays
are to SM gauge-boson pairs, or radiative decays to another v-glueball
and a photon (or perhaps a $Z$ boson.)  We will demonstrate  that
detection should be straightforward, if the mass $M$ of the
quirk $X$ is small enough to give a reasonable cross-section, and
$\Lambda_v$ is large enough to ensure the v-glueballs decay promptly.
Several v-glueballs form di-photon resonances, which should
be easy to detect if their decays are prompt.
Unlike \cite{HVWis}, or especially \cite{hvstudy1}, it appears that
traditional cut-based analysis on ordinary events with jets and
photons will be sufficient.  For displaced decays, however, special
experimental techniques are always needed.  There are a number of
different signatures, and the optimal search strategy is not obvious.

The present paper is organized as follows. In Sec.~2, we introduce our
model and systematically describe the v-sector
operators and the v-glueball states. In Sec.~3, we describe the effective
action coupling the two sectors and the SM matrix elements relevant
for the decays.  Our main results for the decay modes and their
branching fractions appear in Sec.~4.  We conclude in Sec.~5 with some
final comments and perspective.  Additional results appear in the
Appendix.


\section{The model and the hidden valley sector}

\subsection{Description of the Model}\label{model}
 Consider adding to the standard model (SM) a new gauge group $G$,
with a confinement scale $\Lambda_v$ in the 1--1000 GeV range.  We
will refer to this sector as the ``hidden valley'', or the
``v-sector'' following \cite{SZ}.  What makes this particular
confining hidden valley special is that it has no light charged
matter; its only light fields are its gauge bosons, which we will call
``hidden gluons'' or ``v-gluons''.  At low energy, confinement
generates (meta)stable bound states,
``v-glueballs'', from the v-gluons.  The SM is coupled to the hidden
valley sector only through heavy fields $X_r$, in vector-like
representations of both the SM and $G$, with masses of order the TeV
scale.  These states can be produced directly at the LHC, but because
of v-confinement they cannot escape each other; they form a bound
state which relaxes toward the ground state and eventually
annihilates.  The products of the annihilation are often v-glueballs.  (Other
annihilations
lead typically to a hard pair or trio of standard model
particles.)  Thereafter, the v-glueballs decay, giving a potentially
visible signal.

For definiteness, we take the gauge
group $G$ to be $SU(n_v)$, and the particles $X_r$ to transform as a
fundamental representation of $SU(n_v)$ and in complete $SU(5)$
representations of the Standard Model, typically ${\bf 5}+{\bf\bar 5}$ and/or
${\bf 10}+{\bf \overline{10}}$.  We label the fields and their masses
as shown\footnote{In this paper, we normalize hypercharge as
$Y=T_3-Q$, where $T_3$ is the third component of weak isospin.}
in table
\ref{tab reps}.
\begin{table}[h]
\begin{center}
\begin{tabular}[c]{||c||c|c|c|c|c||}\hline
\Trule\Brule Field & $SU(3)$ &$SU(2)$& $U(1)$
& $SU(n_v)$ & Mass
\\ \hline
\Trule\Brule $X_{\bar d}$ & ${\bf 3}$ & ${\bf  1}$  & $\frac13$
&  ${\bf n_v}$  & $m_{\bar d}$
\\ \hline
\Trule\Brule $  X_\ell $ & ${\bf 1}$ & ${\bf 2}$  & ${-\frac12}$
&  ${\bf n_v}$ & $m_{\ell}$
\\ \hline \Trule\Brule
$X_{\bar u} $ & $\bf \bar 3$
& $\bf 1$  & $-\frac23$
& $\bf n_v $ &$m_{\bar u}$
 \\  \hline \Trule\Brule
$X_q$ & $\bf 3$ & $\bf 2$  & $\frac16$
& $\bf n_v$ &$m_q$
\\ \hline \Trule\Brule
$X_e$ & ${\bf 1}$ & ${\bf 1}$  & ${\bf  1}$
&  ${\bf n_v}$ & $m_e$
\\ \hline
\end{tabular}
\end{center}
\caption{\small The new fermions $X_r$ that couple the hidden valley
sector to the SM sector.}
\label{tab reps}
\end{table}
In this paper, we will calculate their effects as a function of $m_r$.
The approximate global $SU(5)$ symmetry of the SM gauge
couplings suggests that the masses $m_{\bar d}$ and $m_\ell$ should be
roughly of the same order of magnitude, and similarly for the masses
$m_q, m_{\bar u}, m_e$.  It is often more convenient to express the
answer as a function of the (partially redundant) dimensionless parameters
\begin{equation}
\label{rho}
\rho_r\equiv m_r/M\ .
 \end{equation}
Here $M$ is a mass scale that can be chosen arbitrarily; depending on parameters, it is usually most natural to take it
to be the mass of the lightest $X_r$ particle.

Integrating out these heavy particles generates an effective
Lagrangian \Leff\ that couples the v-gluons and the SM gauge bosons.
The terms in
the effective Lagrangian are of the form (\ref{loop terms}), with
operators ${\cal O}_v^{(d)}$ constructed from the gauge invariant
combinations\footnote{Here we represent the v-gluon fields as
$\Fc_{\mu\nu}= \Fc_{\mu\nu}^a\,T^a$, where $T^a$ denote the generators
of the $SU(n_v)$ algebra with a common normalization $\tr T^a T^b =
\frac12\, \delta^{ab}.$ } $\tr \Fc_{\mu\nu}\Fc_{\alpha\beta}$ and $\tr
\Fc_{\mu\nu}\Fc_{\alpha\beta} \Fc_{\delta \sigma }$, contracted according
to different irreducible representations of the Lorentz group.

The interactions in the effective action then allow the v-glueballs in
figure~\ref{fig spectrum}, which cannot decay within the v-sector, to decay to final
states containing SM particles and at most one v-glueball.
This is analogous to the way that the Fermi effective theory, which
couples the quark sector to the lepton sector, permits otherwise
stable QCD hadrons to decay weakly to the lepton sector.  As is also
true for leptonic and semileptonic decays of QCD hadrons, our
calculations for v-hadrons decaying into SM particles simplify because
of the factorization of the matrix elements into a purely SM part and
a purely hidden-sector part.  To compute the v-glueball decays, we
will only need the following factorized matrix elements,
involving terms in the effective action of dimension eight:
\begin{equation}
\label{MatEl1}
 \ME{SM}{{{\cal O}_s^{(8-d)}}}{0} \ME{0}{{{\cal O}_v^{(d)}}}{\Theta_\kappa} \ ,
\end{equation}
\begin{equation}
\label{MatEl2}
 \ME{SM}{{{\cal O}_{s}^{(8-d)}}}{0} \ME{\Theta_{\kappa'}}{{{\cal O}_v^{(d)}}}{\Theta_\kappa} \ .
\end{equation}
Here $d$ is the mass dimension of the operator in the v-sector,
$\langle SM|$ schematically represents a state built from Standard
Model particles, and $|\Theta_\kappa\rangle$ and
$|\Theta_{\kappa'}\rangle$ refer to v-glueball states with quantum
numbers $\kappa$, which include spin $J$, parity $P$ and
charge-conjugation $C$.  We will see later that we only need to consider
$d=4$ and $6$; there are no dimension $D=8$ operators in \Leff\ for
which $d=5$, since there are no appropriate dimension-three SM
operators to compensate.  The SM part $\ME{SM}{{{\cal
O}_s^{(8-d)}}}{0}$ can be evaluated by the usual perturbative methods
of quantum field theory, but a computation of the hidden-sector matrix
elements $\ME{0}{{{\cal O}_v^{(d)}}}{\Theta_\kappa}$ and
$\ME{\Theta_{\kappa'}}{{{\cal O}_v^{(d)}}}{\Theta_\kappa}$ requires
the use of non-perturbative methods.

\subsection{Classification of v-glueball states}
In this section we shall classify the nonvanishing v-sector
matrix elements.  A v-glueball state $\Theta_\kappa$ with quantum
numbers $J^{PC}$ can be created by certain operators ${\cal
O}_v^{(d)}$ acting on the vacuum $|\,0\rangle$.  We wish to know which
matrix elements, $\ME{0}{{{\cal
O}_v^{(d)}}}{\Theta_\kappa}$ and $\ME{\Theta_\kappa'}{{{\cal
O}_v^{(d)}}}{\Theta_\kappa}$, are nonvanishing.
This is equivalent to finding how the
operators in various Lorentz representations are projected onto states
with given quantum numbers $J^{PC}$.  Their classification was carried
out in \cite{Jaffe}. At mass dimension $d=4$ there are four different
operators transforming in irreducible representations of the Lorentz
group. These are shown\footnote{As explained in~\cite{Jaffe}, when an operator
${\cal O}_v^{\xi}$ is conserved and the associated symmetry is not
spontaneously broken, some states must decouple. For example,
with
$$
\langle 0|\, T_{\mu\nu}|\,1^{-+}\rangle = (p_\mu\epsilon_\nu+p_\nu\epsilon_\mu){\bf F^T_{1^{-+}}}\ ,
$$
the conservation of $T_{\mu\nu}$ requires ${\bf F^T_{1^{-+}}}=0$, and thus $T$ does not create
a $1^{-+}$ state.  Similarly
$$
 \ME{0}{\,T_{\mu\nu}}{\,0^{++}}=  (a p^2 g_{\mu\nu} +b p_\mu p_\nu){\bf F^T_{0^{++}}},
$$
where $a$ and $b$ are some functions of $p^2$, must vanish for $T_{\mu\nu}$
conserved and traceless.  Note that the trace anomaly complicates this
discussion, but its effect in this model is minimal; see Sec.~3.1 below.}
in table \ref{dim4}. From now on, we denote the
operators ${\cal O}_v^\xi$, where $\xi$ runs over different
irreducible operators $\xi= S, P, T, L, \cdots$.
\begin{table}[h]
\begin{center}
\begin{tabular}[c]{cc}\hline\hline \Trule\Brule
\ Operator ${\cal O}^{\xi}_v$&$J^{PC}$
\\ \hline \Trule
\scriptsize $S= \tr \Fc_{\mu\nu}\Fc^{\mu\nu}$ & $0^{++}$
\\
\scriptsize $P=\tr \Fc_{\mu\nu}\tilde \Fc^{\mu\nu}$ & $0^{-+}$
\\
\scriptsize $T_{\alpha\beta}=\tr  \Fc_{\alpha\lambda}\Fc^{~\lambda}_{\beta} -\frac14\, g_{\alpha\beta} S$ & $2^{++}$, $1^{-+}$, $0^{++}$
\\
 \scriptsize  $ L_{\mu\nu\alpha\beta}=\tr \Fc_{\mu\nu}\Fc_{\alpha\beta} -\frac12 \, (g_{\mu\alpha}T_{\nu\beta}+g_{\nu\beta}T_{\mu\alpha}-g_{\mu\beta}T_{\nu\alpha}-g_{\nu\alpha}T_{\mu\beta})
$ & $2^{++}$, $2^{-+}$ \\ \Brule
\scriptsize $-\frac{1}{12}\,(g_{\mu\alpha}g_{\nu\beta}-g_{\mu\beta}g_{\nu\alpha})S +\frac{1}{12}\,\epsilon_{\mu\nu\alpha\beta}P$   &
\\ \hline\hline
\end{tabular}
\end{center}
\caption{\small The dimension $d=4$ operators, and the states that can
be created by these operators \cite{Jaffe}. We denote $\tilde
\Fc_{\mu\nu}=\frac12\, \epsilon_{\mu\nu\alpha\beta}\Fc^{\alpha\beta}$.}
\label{dim4}
\end{table}

The study of irreducible representations of dimension-six operators is
more involved. A complete analysis in terms of electric and magnetic
gluon fields, $\vec E_a$ and $\vec B_a$, was also presented in
\cite{Jaffe}, with a detailed description of the operators and the
states contained in their spectrum.
There are only two such operators of relevance for our work,
which we denote $\Omega^{(1)}_{\mu\nu}$ and
$\Omega^{(2)}_{\mu\nu}$ as shown in table \ref{dim6}. The other dimension-six operators simply cannot be combined with any SM operator to make a dimension-eight interaction.
\begin{table}[ht]
\begin{center}
\begin{tabular}[c]{cc}\hline\hline\Trule\Brule
\ Operator ${\cal O}^{\xi}_v$ &$J^{PC}$
\\ \hline \Trule
\scriptsize $\Omega^{(1)}_{\mu\nu}= \tr \Fc_{\mu\nu}\Fc_{\alpha\beta}\Fc^{\alpha\beta}$ & $1^{--}$, $1^{+-}$ \\
\Brule \scriptsize $\Omega^{(2)}_{\mu\nu}= \tr \Fc_\mu^{\alpha} \Fc^{\beta}_{\alpha}\Fc_{\beta\nu}$& $1^{--}$, $1^{+-}$
\\ \hline\hline
\end{tabular}
\end{center}
\caption{\small The important $d=6$ operators. 
The states that can be created by these operators are shown \cite{Jaffe}.}
\label{dim6}
\end{table}


\subsection{Matrix elements}

As we saw, the matrix elements are factorized into
a purely SM part and a purely v-sector part.
We will first
consider the v-sector matrix elements relevant to v-glueball
transitions,
$\ME{0}{{{\cal
O}_v^{\xi}}}{\Theta_\kappa}$ and $\ME{\Theta_{\kappa'}}{{{\cal
O}_v^{\xi}}}{\Theta_\kappa}$, where $|\Theta_\kappa\rangle$ and
$|\Theta_{\kappa'}\rangle$ refer to v-glueball states with given
quantum numbers and ${\cal O}^{\xi}_v$ is any of the operators in
tables~\ref{dim4} and~\ref{dim6}.

It is convenient to write the most general possible matrix element in
terms of a few Lorentz invariant amplitudes or form factors. For the annihilation matrix elements we will write
\begin{equation}
\label{HVme}
 \ME{0}{{{\cal O}_v^{\xi}}}{\Theta_\kappa}= \Pi_{\kappa,
 \mu\nu\cdots}^\xi {\bf F}_\kappa^\xi \ ,
\end{equation}
where ${\bf F_\kappa^\xi}$ is the decay constant of the v-glueball
$\Theta_\kappa$, and $\Pi_{\kappa, \mu\nu\cdots}^\xi$ is determined by
the Lorentz representations of $\Theta_\kappa$ and ${\cal O}_v^{\xi}$. In table~\ref{constant_decay} we list
$\Pi_{\kappa, \mu\nu\cdots}^\xi$ for each operator.

The decay constants ${\bf F}_\kappa^\xi $ depend on the internal
structure of the v-glueball states and, with the exception of those 
that vanish due to conservation laws (see footnote 6), must be determined by
non-perturbative methods, for instance, by numerical calculations in
lattice gauge theory.  Only the first three non-vanishing
decay constants in table~\ref{constant_decay} have been calculated,
for  $SU(3)$ Yang-Mills theory \cite{Morningstar2}, although
the reported values are not expressed in a continuum renormalization
scheme.  The other decay constants have not been computed.

Likewise, the transition matrix elements $\ME{\Theta_{\kappa'}}{{{\cal O}_v^{\xi}}}{\Theta_\kappa}$ are of the form
\begin{equation}
\label{HVtm}
 \ME{\Theta_{\kappa'}}{{{\cal O}_v^{\xi}}}{\Theta_\kappa}= \Pi_{\kappa\kappa', \mu\nu\cdots}^\xi {\bf M}_{\kappa, \kappa'}^\xi,
\end{equation}
where now ${\bf M}_{\kappa, \kappa'}^\xi$ is the transition matrix, which depends only on the transferred momentum. In
table~\ref{transition_matrix} we have listed $\Pi_{\kappa\kappa', \mu\nu\cdots}^\xi$ for the simplest cases considered later in this
work. In several other cases more than one Lorentz structure $\Pi_{\kappa\kappa', \mu\nu\cdots}^\xi$ contributes to the transition element. In such cases, since none of these matrix elements are known from numerical simulation, 
we will usually simplify the problem by using the lowest partial-wave approximation for the amplitudes. More details will follow in Sec.~\ref{decay rates}.

\begin{table}[ht]
\begin{center}
\begin{tabular}[c]{ccc}\hline\hline \Trule\Brule
${\cal O}^{\xi}_v$  ($\Theta_\kappa$) & $\Pi_{\kappa, \mu\nu\cdots}^\xi$ & $ {\bf F}_\kappa^\xi$
\\ \hline\Trule
\scriptsize $S$ $(0^{++})$ &  $\mathbf{1}$  &\small$  \mathbf{F^S_{0^{++}}}$ \\
\scriptsize $P$ $(0^{-+})$ &  $\mathbf{1}$ &\small$  \mathbf{F^P_{0^{-+}}}$ \\
\scriptsize $T_{\alpha\beta}$ $(0^{++})$ &  $g_{\alpha\beta}-\frac{p_\alpha p_\beta}{p^2}$  & 0 \\
\scriptsize $T_{\alpha\beta}$ $(1^{-+})$ &  $p_\alpha \epsilon_\beta + p_\beta \epsilon_\alpha$ & 0 \\
\scriptsize $T_{\alpha\beta}$ $(2^{++})$ &  $\epsilon_{\alpha\beta}$ &\small$ \mathbf{F^T_{2^{++}}}$ \\
\scriptsize $L_{\mu\nu\alpha\beta}$ $(2^{++})$ & \scriptsize $\epsilon_{\mu\alpha} \mathcal{P}_{\nu\beta}+\epsilon_{\nu\beta} \mathcal{P}_{\mu\alpha}-\epsilon_{\nu\alpha} \mathcal{P}_{\mu\beta}-\epsilon_{\mu\beta} \mathcal{P}_{\nu\alpha}$  & $\mathbf{F^L_{2^{++}}}$ \\
\scriptsize $L_{\mu\nu\alpha\beta}$ $(2^{-+})$ & \scriptsize $(\epsilon_{\mu\nu\rho\sigma}\epsilon^\sigma_{~\beta}p^\rho
p_\alpha - \epsilon_{\mu\nu\rho\sigma}\epsilon^\sigma_{~\alpha}p^\rho
p_\beta
$ & $\mathbf{F^L_{2^{-+}}}$\\
 &\scriptsize  $+ \epsilon_{\alpha\beta\rho\sigma}\epsilon^\sigma_{~\nu}p^\rho p_\mu -
\epsilon_{\alpha\beta\rho\sigma}\epsilon^\sigma_{~\mu}p^\rho p_\nu)/p^2$  &\\
\scriptsize $\Omega^{(n)}_{\mu\nu}$ $(1^{--})$ &  $m_{1^-}(p_\mu \epsilon_\nu - p_\nu \epsilon_\mu)$ & $\mathbf{F^{\Omega^{(n)}}_{1^{--}}}$\\
\Brule \scriptsize $\Omega^{(n)}_{\mu\nu}$ $(1^{+-})$ &  $m_{1^+}\epsilon_{\mu\nu\alpha\beta}(p^\alpha \epsilon^\beta - p^\beta \epsilon^\alpha)$ & $\mathbf{F^{\Omega^{(n)}}_{1^{+-}}}$\\
\hline\hline
\end{tabular}
\end{center}
\caption{\small  Annihilation matrix elements. $\epsilon_\mu$ and $\epsilon_{\mu\nu}$ are the polarization vectors of $1^{--}, 1^{+-}$ and polarization tensor of $2^{++},2^{-+}$ respectively. ${\mathcal{P}}_{\alpha\beta}= g_{\alpha\beta} - 2 p_\alpha p_\beta/p^2$. $m_{1^-},m_{1^+}$ are the masses of the $1^{--}, 1^{+-}$ states; their appearance merely reflects our normalization convention.
}
\label{constant_decay}
\end{table}

\begin{table}[ht]
\begin{center}
\begin{tabular}[c]{ccc}\hline\hline\Trule\Brule
 ${\cal O}^{\xi}_v$  ($\Theta_\kappa$ $\Theta_{\kappa'}$) & $\Pi_{\kappa \kappa', \mu\nu\cdots}^\xi$ & $ {\bf M}_{\kappa\kappa'}^\xi$
\\ \hline \Trule

\scriptsize $P$ $(0^{-+},0^{++})$ &  $\mathbf{1}$ & $\mathbf{M_{0^+0^-}^P}$\\
\scriptsize $\Omega_{\mu\nu}^{(n)}$ $(1^{--}, 0^{++})$ &  $p_\mu \epsilon_\nu - p_\nu \epsilon_\mu $ & $\mathbf{M_{1^{--}0^{++}}^{\Omega^{(n)}}}$\\
\scriptsize $\Omega_{\mu\nu}^{(n)}$ $(1^{+-}, 0^{-+})$ &  $p_\mu \epsilon_\nu - p_\nu \epsilon_\mu$ & $\mathbf{M_{1^{+-}0^{-+}}^{\Omega^{(n)}}}$\\

\scriptsize $\Omega_{\mu\nu}^{(n)}$ $(1^{--},0^{-+})$ &  $\epsilon_{\mu\nu\alpha\beta}p^\alpha \epsilon^\beta  $ & $\mathbf{M_{1^{--}0^{-+}}^{\Omega^{(n)}}}$\\
\scriptsize $\Omega_{\mu\nu}^{(n)}$ $(1^{+-},0^{++})$ &  $\epsilon_{\mu\nu\alpha\beta}p^\alpha \epsilon^\beta  $ & $\mathbf{M_{1^{+-}0^{++}}^{\Omega^{(n)}}}$ \\ \hline
\Trule \scriptsize $P$ $(1^{--},1^{+-})$ &  $\epsilon^+ \cdot \epsilon^-$ &  $\mathbf{
M_{1^{--}1^{+-}}^P}$\\
 \scriptsize $L$ $(1^{--},1^{+-})$ &  $\epsilon_{\mu\nu\rho\sigma}p^\rho {\epsilon^-}^\sigma(p_\alpha{\epsilon^+}_\beta - p_\beta{\epsilon^+}_\alpha )
 + \mu\nu \leftrightarrow \alpha\beta - {\rm traces} $ &  $\mathbf{
M_{1^{--}1^{+-}}^L}$\\

\Brule \scriptsize $\Omega_{\mu\nu}^{(n)}$ $(2^{-+},1^{+-})$ &  $p_\mu \epsilon_{\nu\alpha}\epsilon^\alpha - p_\nu \epsilon_{\mu\alpha}\epsilon^\alpha$ & $\mathbf{\bf M_{2^{-+}1^{+-}}^{\Omega^{(n)}}}$\\
\Brule \scriptsize $\Omega_{\mu\nu}^{(n)}$ $(1^{+-},2^{++})$ &  $ \epsilon_{\mu\nu\alpha\beta} \epsilon^{\alpha\lambda} \tilde \epsilon_\lambda p^\beta  $, $ \epsilon_{\mu\nu\alpha\beta} \epsilon^{\alpha\lambda}  p_\lambda \tilde \epsilon^\beta $ & $\mathbf{\bf M_{1^{+-}2^{++}}^{\Omega^{(n)}}}$
\\ \hline\hline
\end{tabular}
\end{center}
\caption{\small Transition matrix elements. Momentum of the final glueball $\Theta_{\kappa'}$ is denoted $p^\mu$; $\epsilon^{\alpha}$ and $\epsilon^{\alpha\beta}$ are polarization tensors of spin~1 and spin~2 states respectively. The bottom part of the table contains matrix elements in the lowest partial wave approximation.}
\label{transition_matrix}
\end{table}

Clearly, any numerical results arising from our formulas, as we
ourselves will obtain in our LHC study \cite{vglue2}, will be subject
to some large uncertainties, due to the unknown matrix elements.  Of
course, with sufficient motivation, such as a hint of a discovery,
many of these could be determined through additional lattice gauge
theory computations.

Now we turn to the SM part of the matrix element, which can be treated
perturbatively, since we will only consider v-glueballs with
masses well above $\Lambda_{QCD}$.\footnote{We will do all our
calculations at SM-tree level; loop corrections
for v-glueball decays to ordinary gluons should be accounted
for when precision is required.}  In all of our
calculations, the SM gauge-boson field-strength tensors, which appear
in the operators, are replaced in the matrix element by the
substitution $G_{\mu\nu}\leftrightarrow k_\mu\varepsilon_\nu-
k_\nu\varepsilon_\mu$. For example, for a transition to two gauge
bosons, we write\footnote{Note that one has to take into account a factor of 2   which comes from the two different ways of contracting each $G_{ \mu\nu}^a$ operator with $|k_1,\varepsilon^a_1;k_2,\varepsilon^b_2\rangle $. This factor then cancels an explicit $\frac12$ factor appearing in the normalization of the trace.  }

\begin{equation}
 \ME{k_1,\varepsilon^a_1;k_2,\varepsilon^b_2}{{\tr
G_{\mu\nu}G_{\alpha\beta}}}{0} =\delta^{ab}
(k^1_\mu\varepsilon^1_\nu-
k^1_\nu\varepsilon^1_\mu)(k^2_\alpha\varepsilon^2_\beta-
k^1_\alpha\varepsilon^2_\beta),
\end{equation}
where $k^{1(2)}$, $\varepsilon^{1(2)}$ are the gauge-bosons' momenta and
polarizations respectively. Later in the text we will sometimes use the
following notation for the SM matrix elements
\begin{equation}\label{SMme}
 \ME{SM}{{{\cal O}_s^{\eta}}}{0}= h^{\mu\nu\cdots}_\eta\ ,
\end{equation}
where  $h^{\mu\nu\cdots}_\eta=h^{\mu\nu\cdots}_\eta(k_1,k_2,\cdots)$ is a function of the momenta of the SM particles in the final state.



\section{Effective Lagrangian}
\label{Effective lagrangian}
In this section we discuss the effective action \Leff\ linking the SM
sector with the v-sector, and discuss the general form of the
amplitudes controlling v-glueball decays.  We will confirm that all
the important decay modes are controlled by $D=8$ operators involving the
$d=4$ and 6 operators listed in tables \ref{dim4} and \ref{dim6}.

\subsection{Heavy particles and the computation of ${\cal L}_{{\rm eff}}$}
\begin{figure}[htb]
\begin{minipage}[b]{0.5\linewidth} 
\centering
\includegraphics[width=6cm]{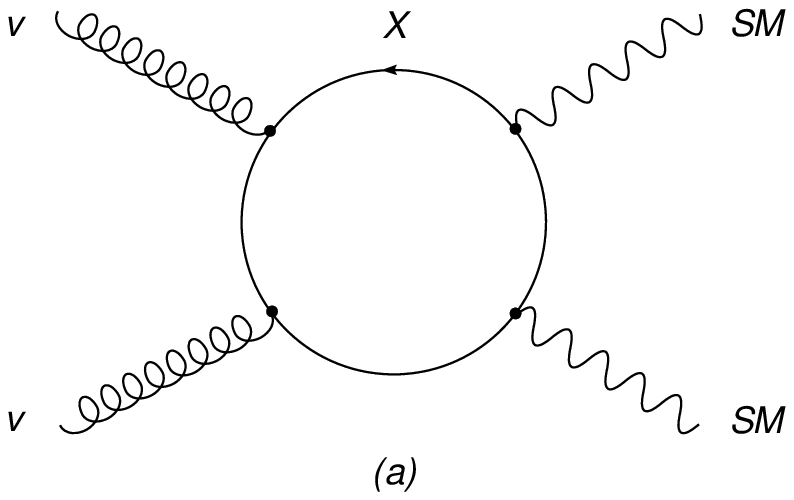}
\end{minipage}
\hspace{0.0cm} 
\begin{minipage}[b]{0.5\linewidth}
\centering
\includegraphics[width=6cm]{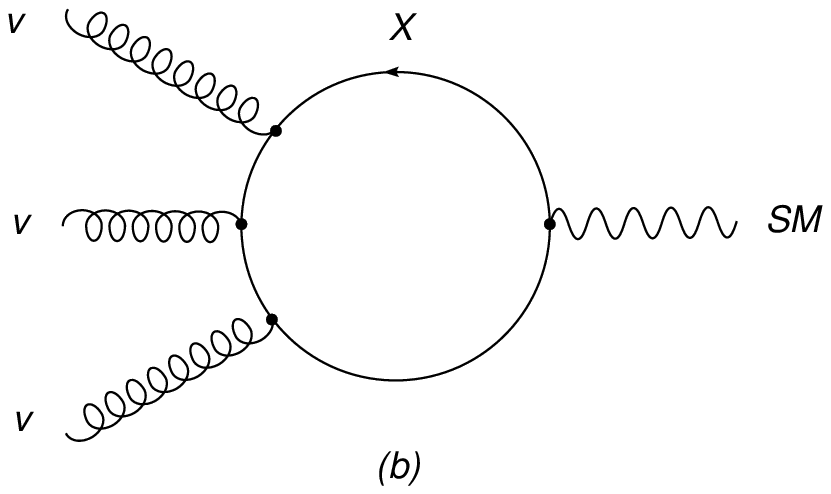}
\end{minipage}
\caption{\small Diagrams contributing to the effective action}
\label{diagrams fig}
\end{figure}

The low-energy interaction of v-gluons and v-glueballs with SM
particles is induced through a loop of heavy $X$-particles. In this
section we present the one-loop effective Lagrangian that describes
this interaction, to leading non-vanishing order in $1/M$, namely
$1/M^4$, which we will see is sufficient for inducing all v-glueball
decays.  The relevant diagrams all have four external gauge boson
lines, as depicted in figure \ref{diagrams fig}.  They give the
amplitude for scattering of two v-gluons to two SM gauge bosons, of
either strong (gluons $g$), weak ($W$ and $Z$) or hypercharge (photon
$\gamma$ or $Z$) interactions (figure \ref{diagrams fig}a), as
well as the
conversion of three v-gluons to a $\gamma$ or $Z$ (figure
\ref{diagrams fig}b).

The dimension-eight operators appearing in the action can be found in
studies of Euler-Heisenberg-like Lagrangians in the literature. Within
the SM, effective two gluon - two photon, four gluon, and
three gluon - photon vertices can be found in \cite{Shifman
charmonium}, \cite{Strassler:1992nc} and \cite{Groote}
respectively. These results can be adapted for our
present purposes.

We introduce now some notation, defining $G_{\mu\nu}^1\equiv
B_{\mu\nu}$, $G_{\mu\nu}^2\equiv F_{\mu\nu}$ and $G_{\mu\nu}^3\equiv
G_{\mu\nu}$, which are the field tensors of the $U(1)_Y$, $SU(2)$ and
$SU(3)$ SM gauge groups. We denote their couplings $g_{i}$, $i=1,2,3$,
while $g_v$ is the coupling of the new group $SU(n_v)$. In terms of
the operators from tables~\ref{dim4} and~\ref{dim6}, the effective
Lagrangian reads

\begin{multline}
\label{eff_lag full2}
 {\cal L}_{{\rm eff}}= \frac{g_v^2}{(4\pi)^2M^4}\left[g_1^2\chi_1 {B}^{\mu\nu}{B}^{\rho\sigma} +
g_2^2\chi_2 \tr{F}^{\mu\nu}{F}^{\rho\sigma} +
g_3^2\chi_3 \tr{G}^{\mu\nu}{G}^{\rho\sigma} \right]\\
 \ \ \ \ \ \ \ \ \ \ \ \ \ \ \ \ \ \ \ \ \times
\left(
  \frac{1}{60} \,S\, g_{\mu\rho}g_{\nu\sigma}
  + \frac{1}{45} \,P\, \epsilon_{\mu\nu\rho\sigma}
 + \frac{11}{45} \,T_{\mu\rho}g_{\nu\sigma}\, - \frac{1}{30}  \,L_{\mu\nu\rho\sigma}\,  \right) \\ +
 \frac{g_v^3g_1}{(4\pi)^2M^4}\,\chi\left(\frac{14}{45}\,
 B^{\mu\nu}\Omega^{(1)}_{\mu\nu} - \frac{1}{9}\,
 B^{\mu\nu}\Omega^{(2)}_{\mu\nu}\right).
\end{multline}
The coefficients $\chi_i$ and $\chi$ encode the masses of the heavy
particles from table~\ref{tab reps} and their couplings to the SM gauge
groups. They are summarized in table~\ref{xi}.

\begin{table}[ht]\label{ratios}
\begin{center}
 \begin{tabular}[c]{||c||c||}\hline \Trule\Brule
\  &$\chi$ , $\chi_i$
\\ \hline \Trule\Brule
$\chi_1$ & $ \frac{1}{ 3 \rho_{\bar d}^4}+\frac{1}{2 \rho_{ l}^4} +\frac{4}{3 \rho_{ \bar u}^4}+\frac{1}{6 \rho_{q}^4}+\frac{1}{\rho_e^4} $
\\ \hline \Trule\Brule
$\chi_2$ & $\frac{1}{ \rho_{l}^4}+\frac{3}{ \rho_q^4} $
\\ \hline \Trule\Brule
$\chi_3$ & $\frac{1}{\rho_{\bar d }^4}+\frac{1}{\rho_{\bar u}^4} +\frac{2}{\rho_q^4} $
\\ \hline \Trule\Brule
$\chi$ & $\frac{1}{\rho_{\bar d}^4}-\frac{1}{\rho_{l}^4} -\frac{2}{\rho_{\bar u}^4}+\frac{1}{\rho_q^4}+\frac{1}{\rho_e^4}$
\\ \hline
\end{tabular}
\end{center}
\caption{\small The coefficients $\chi$ arise from a sum over the SM charges of $X$ particles running in the loop. The $\chi_i$, $i=1,2,3$, arise from the diagram in figure~\ref{diagrams fig}(a) with two
external SM gauge bosons of group $i$, while $\chi$ is determined by the
diagram~\ref{diagrams fig}(b) with a single hypercharge-boson on an external line.
The $\rho_r$ are defined in (\ref{rho}).}
\label{xi}
\end{table}

The effective Lagrangian (\ref{eff_lag full2}) can be compactly written as
\begin{equation}\label{eff_lag}
 {\cal L}_{{\rm eff}} =\sum^3_{i=1}\sum_{\xi} \frac{g_v^{\frac{d_{\xi}}{2}}g_i^{4-\frac{d_{\xi}}{2}}}{(4\pi)^2 M^4} \,\Xi^{i}_{\xi} {\cal O}_{s}^{\eta(\xi,i)}\cdot {\cal O}_{v}^{\xi},
\end{equation}
where the sum is over operators and different ways to contract Lorentz
indices. The notation $\eta(\xi,i)$ is to make explicit that for each
$\xi$ and $i$ there is at most one SM operator ${\cal O}_{s}^{\eta}$
multiplying ${\cal O}_{v}^{\xi}$ in the effective Lagrangian (see
table~\ref {cs}).

The mass dimension of ${\cal O}_{v}^{\xi}$ is denoted $d_{\xi}$,
and the $\Xi^{i}_{\xi}$ are dimensionless coefficients given by
\be
\label{xi Big}
\Xi^{i}_{\xi} = \left\{ \begin{array}{ll}
 \chi^i C_{\xi} & d_{\xi}= 4\\
 \chi C_{\xi} & d_{\xi} = 6 \  .
  \end{array} \right.
\ee
The $C_{\xi}$ are coefficients that depend only on the v-sector
operators and the SM operator with which they are
contracted; they are also given in table~\ref{cs}.

These values for the $C_{\xi}$ are valid around the scale $M$,
and they will be altered by perturbative renormalization between
this scale and a lower scale closer to the glueball masses,
at which the nonperturbative
matrix elements are evaluated.\footnote{Decays of v-glueballs to
standard model gauge bosons are affected by the trace anomaly,
but minimally, because both sectors' trace anomalies
must be non-zero, and that of the SM is small
at the scale of the v-glueball masses.}
These renormalization effects
(which will impact v-glueball lifetimes
but cancel out of most
branching fractions) can be computed, but are only useful to discuss
once one has concrete values for the decay constants and matrix elements
in a definite renormalization scheme, which at present is not 
available.  We will not discuss them further here.

\begin{table}[ht]
\begin{center}
\begin{tabular}[c]{ccc|ccc} \hline\hline \Trule\Brule
  \small ${\cal O}_v^{\xi}$ &\small $C_{\xi}$&\small
$ {\cal O}_s^\eta \cdot {\cal O}_v^{\xi}$ &
\small ${\cal O}_v^{\xi}$ &\small
  $C_{\xi}$&\small $ {\cal O}_s^\eta \cdot {\cal O}_v^{\xi}$
\\ \hline \Trule $S$ & $\frac{1}{60} $& $(\tr G^i_{\mu\nu}G^{i\mu\nu})\ S$
& $T$ & $\frac{11}{45}$ & $
(\tr G^{i\mu\lambda}
G^{i\nu}_{~\lambda}) \ T_{\mu\nu} $\\
$P$ & $\frac{2}{45}$ & $(\tr G^i_{\mu\nu}\tilde G^{i\mu\nu}) \ P$ &
$\Omega^{(1)}$&$\frac{14}{45}$&$G^{1\mu\nu}\ \Omega^{(1)}_{\mu\nu} $\\
\Brule $L$ & $-\frac{1}{30}$ & $(\tr
G^{i\mu\nu}G^{i\alpha\beta})\ L_{\mu\nu\alpha\beta} $ &
$\Omega^{(2)}$&$-\frac{1}{9}$&$G^{1\mu\nu}  \ \Omega^{(2)}_{\mu\nu} $\\

\hline\hline
\end{tabular}
\end{center}
\caption{\small List of coefficients $C_{\xi}$ and contractions of the operators ${\cal O}_v^{\xi}$ introduced in tables~\ref{dim4} and~\ref{dim6}. $G^i_{\mu\nu}$ represents the field-strength tensor of the $i^{th}$ SM group.}
\label{cs}
\end{table}

The coefficients $\chi$ and $\chi_i$ in table \ref{xi} determine the
relative coupling of v-gluons to the electroweak-sector gauge
bosons $W^i_\mu$ and $B_\mu$ for the $SU(2)$ and $U(1)_Y$ factors
respectively. For applications it is convenient to convert these to
the couplings to the photons $\gamma$, $W$ and $Z$ bosons. 
We introduce the following coefficients
\be
\label{chis}
\chi_{\gamma}\equiv \chi_1 +\chi_2 /2 ,\qquad \chi_Z\equiv  \frac{ \sin^4 \theta_W \chi_1 +\cos^4 \theta_W \chi_2 /2}{ \cos^2 \theta_W},\\
\ds \chi_W \equiv  \chi_2, \qquad \chi_{\gamma Z} \equiv \frac{\cos^2 {\theta_W} \chi_2 - 2\sin^2{\theta_W} \chi_1}{\cos\theta_W}, \qquad \chi_s\equiv \chi_3 \ ,
\ee
where $\theta_W$ is the weak mixing angle.
We will often use these coefficients instead of $\chi_i$ in the effective
Lagrangian (\ref{eff_lag full2}), with a corresponding substitution of
field tensors and couplings.


\subsection{Decay amplitudes }
\label{amplitsect}

Now, using (\ref{HVme}), (\ref{SMme}) and the couplings from
(\ref{eff_lag}), we obtain that the amplitude for a decay of a
v-glueball into SM particles is given by
\begin{multline}
\label{amplitude1}
 {\cal M} = \frac{g_v^{\frac{d_\xi}{2}}g_i^{4-\frac{d_\xi}{2}}}{(4\pi)^2 M^4} \,\Xi_\xi^i(\rho_{\bar u},...,\rho_e)\ME{SM}{{{\cal O}_s^{\eta}}}{\,0} \ME{0}{{{\cal O}_v^{\xi}}}{\,\Theta_\kappa}= \\
=\frac{g_v^{\frac{d_\xi}{2}}g_i^{4-\frac{d_\xi}{2}}}{(4\pi)^2 M^4} \Xi_\xi^i(\rho_{\bar u},...,\rho_e)  f^i_{\xi,\eta}(p,q_1,q_2,...) \mathbf{F}_\kappa^\xi,
\end{multline}
 where
 $$f^i_{\xi,\,\eta}(p,k_1,k_2,...) = h^{\mu\nu\cdots}_\eta(k_1,k_2,...)
 \Pi_{\kappa,\,\mu\nu\cdots}^\xi(p) $$
 encodes all the information about
 the matrix element that can be determined from purely perturbative
 computations and Lorentz or gauge invariance, and
 $\mathbf{F}_\kappa^\xi$ is the v-glueball decay
 constant.  See Eq.~(\ref{xi Big}) for the definition of $\Xi$ and
 Eq.~(\ref{rho}) and table \ref{xi} for the definition of $\rho$.

Similarly, using (\ref{HVtm}), (\ref{SMme}) and (\ref{eff_lag}), the
amplitude for the decay of a v-glueball into another v-glueball and
SM particles reads
\begin{multline}\label{amplitude2}
{\cal M} = \frac{g_v^{\frac{d_\xi}{2}}g_i^{4-\frac{d_\xi}{2}}}{(4\pi)^2 M^4} \,\Xi_\xi^i(\rho_{\bar u},...,\rho_e) \ME{SM}{{{\cal O}_s^{\eta}}}{\,0}  \ME{\Theta_{\kappa'}}{{{\cal O}_v^{\xi}}}{\,\Theta_\kappa} =
\\ =\frac{g_v^{\frac{d_\xi}{2}}g_i^{4-\frac{d_\xi}{2}}}{(4\pi)^2 M^4} \,\Xi_\xi^i(\rho_{\bar u},...,\rho_e)  f^i_{\kappa\kappa';\xi,\eta}(p,p',k_1,k_2,...) \mathbf{M}_{\kappa\kappa'}^\xi (k) .
\end{multline}
Here $\mathbf{M}_{\kappa\kappa'}^\xi(k)$ is the glueball-glueball transition matrix, which for given masses of $\Theta_\kappa$ and $\Theta_{\kappa'}$ is a function of transferred momentum $k\equiv p'-p$, and
$$f^i_{\kappa\kappa';\,\xi,\eta}= \Pi_{\kappa \kappa',\,\mu\nu\cdots}^\xi(p,p') h^{\mu\nu\cdots}_\eta(k_1,k_2,...).$$


\section{Decay rates for lightest v-glueballs}
\label{decay rates}

In this section we will compute the decay rates for some of the
v-glueballs in figure~\ref{fig spectrum}.  Let us make a quick summary
of the results to come.

The operators shown in tables \ref{dim4} and \ref{dim6} induce the
dominant  decay modes of the v-glueball states appearing in
figure \ref{fig spectrum}. In the $PC=$++ sector, the lightest
$0^{++}$ and $2^{++}$ v-glueballs will mostly decay directly to pairs
of SM gauge bosons via $S$, $T$ and $L$ operators.  Three-body decays
$2^{++}\rightarrow 0^{++}$ plus two SM gauge bosons are also possible,
but are strongly suppressed by phase space. In the $PC=-$+ sector the
lightest states are the $0^{-+}$ and $2^{-+}$ v-glueballs. These will
also decay predominantly to SM gauge boson pairs, via $P$ and $L$
operators respectively. There are also $C$-changing $2^{-+}\rightarrow
1^{+-}+\gamma$ decays, induced by the $d=6$ $D=8$ operators
$\Omega_{\mu\nu}$ (table \ref{dim6}), but the small mass-splitting
found in the lattice computations \cite{Morningstar} suggests these
decays are probably very rare or absent. In the $PC=$+$-$ sector, the
leading decays are two-body $C$-changing processes, because
$C$-conservation forbids annihilation to pairs of gauge bosons, and
because three-body decays are phase-space suppressed. In
particular, the $1^{+-}$, the lightest v-glueball in that sector, will
decay to the lighter $C$-even states $0^{++}$, $2^{++}$ and $0^{-+}$
by radiating a photon (or $Z$ when it is possible kinematically).  The
same is true for the states in the $PC=--$ sector, with an exception
that the lightest $1^{--}$ v-glueball can annihilate to a pair of SM
fermions through an off-shell photon or $Z$. The latter decay is also
induced by $\Omega^{\mu\nu}$ operators.

We shall study decays of the $0^{++}$, $2^{++}$, $0^{-+}$, $2^{-+}$,
$1^{+-}$ and $1^{--}$ v-glueballs in some detail. Since for this set
of v-glueballs the combination of $J$ and $P$ quantum numbers is
unique, we shall often omit the $C$ quantum number from our formulas
to keep them a bit shorter, referring simply to the $0^+$, $2^+$,
$0^-$, $2^-$, $1^+$ and $1^-$ states.  At the end we shall make some
brief comments about the other states, the
$3^{++}$, $3^{+-}$, $3^{--}$, $2^{+-}$, $2^{--}$ and $0^{+-}$.

Of course the allowed decays and the corresponding lifetimes are
dependent upon the masses of the v-glueballs.  While the results of
Morningstar and Peardon~\cite{Morningstar}, understood as
dimensionless in units of the confinement scale $\Lambda$, can be
applied to any pure $SU(3)$ gauge sector, the glueball spectrum for
$SU(4)$ or $SU(7)$ are not known.  Fortunately, at least for
$SU(n_v)$, the spectrum is expected to be largely independent of
$n_v$.  Still, the precise masses will certainly be different for
$n_v>3$, and for some v-glueballs
this could have a substantive effect
on their lifetimes and branching fractions.

For other gauge groups, however, the spectrum may be qualitatively
different; in particular, the $C$-odd sector may be absent or heavy.
We will briefly discuss this in our concluding section.  The $0^{\pm
+}$ and $2^{\pm+}$ states are expected to be present in any pure-gauge
theory, with similar production and decay channels, and as such are
the most model-independent.  Fortunately, it turns out they are also
the easiest to study theoretically, and, as we will see below and in
our LHC study \cite{vglue2}, the easiest to observe.

\subsection{Light $C$-even sector decays}
We begin with the $C$-even $0^{++}$, $2^{++}$, $0^{-+}$ and $2^{-+}$
v-glueballs, which can be created by dimension 4 operators. The first
three have been studied in some detail in various contexts; see for
example~\cite{Morningstar2,Shifman scalar,Shifman
pseudoscalar,MITbagmodel,glueballs,bagmodel,Loan} and a recent review~\cite{glueballreview}.  The dominant decays
of these states are annihilations $\Theta_\kappa\to G^aG^b$, where
$\Theta_\kappa$ denotes a v-glueball state and $G^a$, $G^b$ is a pair
of SM gauge bosons: $gg$, $\gamma\gamma$, $ZZ$, $W^+W^-$ or $\gamma
Z$.  We will also consider radiative decays $\Theta_{\kappa}\to
\Theta_{\kappa^\prime} +\gamma/Z$, and three-body decays of the form
$\Theta_\kappa\to G^aG^b\Theta_\kappa'$, and will see they are
generally subleading for these states.

Annihilations are mediated by the dimension $d=4$ operators in
Eq.~(\ref{MatEl1}). In particular, we know from the previous discussion
(see \cite{Jaffe} and table \ref{dim4} above) that the $0^{++}$
v-glueball can be annihilated (created) by the operator $S$. The
$0^{-+}$ and $2^{-+}$ states are annihilated by the operators $P$ and
$L_{\mu\nu\alpha\beta}$ respectively. The tensor $2^{++}$ can be
destroyed by both $T_{\mu\nu}$ and $L_{\mu\nu\alpha\beta}$.

Radiative two-body decays are induced by the dimension $d=6$ operators in
Eq.~(\ref{MatEl2}).  However, the decays $\Theta_\kappa\to
\Theta_{\kappa^\prime} +\gamma/Z$ are forbidden if $\Theta_\kappa$ and
$\Theta_{\kappa^\prime}$ are both from the $C$-even subsector.  For
the spectrum in figure~\ref{fig spectrum}, appropriate for $n_v=3$,
the only kinematically allowed radiative decay is therefore $2^{-+}\to
1^{+-}+\gamma$; the $1^{+-}+Z$ final state is kinematically allowed
only for very large $\Lambda_v$.  For $n_v>3$, the glueball spectrum
is believed to be quite similar to $n_v=3$, but the close spacing
between these two states implies that the ordering of masses might be
altered, so that even this decay might be absent for larger~$n_v$.

\paragraph{Decays of the $0^{++}$ state.} The scalar state can
be created or destroyed by the operator $S$.

Then, according to a
general discussion in Sec.~\ref{Effective lagrangian}, the amplitude of the decay of the
scalar to two SM gauge bosons $G^a$ and $G^b$ is given by the expression
\be
\label{2boson decay}
\frac{\alpha_i\alpha_v}{M^4}\,\chi_i\, C_{S}\langle
G^a,G^b|\tr{G}_{\mu\nu}{G}^{\mu\nu}|\,0\rangle\,\langle\, 0| S|\, 0^{++}\rangle,
\ee
where $\alpha_i$ and $\chi_i$ encode the couplings of the bosons $a$ and
$b$ of a SM gauge group $i$ to the loop, introduced in Sec.~\ref{Effective lagrangian}; see (\ref{eff_lag}), (\ref{xi Big}) and table~\ref{xi}.

For the decay of the scalar to two gluons,
(\ref{2boson decay}) takes
the form
\begin{multline}
\label{s decay}
\frac{\alpha_s\alpha_v}{M^4}\chi_s\, C_{S}\langle
g_1^ag_2^b|\,\tr{G}_{\mu\nu}{G}^{\mu\nu}|\,0\rangle\,\langle\, 0| \,S|\,
0^{++}\rangle  =
\\ = \frac{\alpha_s\alpha_v}{M^4}\,\frac{\delta^{ab}}{2}\,\chi_s
C_{S}{\bf F_{0^{++}}^S} 2(k^1_\mu\varepsilon^1_\nu -
k^1_\nu\varepsilon^1_\mu)({k^2}^\mu{\varepsilon^2}^\nu-{k^2}^\nu{\varepsilon^2}^\mu),
\end{multline}
where, according to our conventions, constant $\bf{F_{0^{++}}^S}$
denotes the matrix element $\langle 0| S|\, 0^{++}\rangle$.
We are using the notation
$\alpha_s\equiv \alpha_3, \chi_s\equiv \chi_3$.  The rate of
the decay (accounting for a $1/2$ from Bose statistics) is then given by
\be
\Gamma_{0^{+}\rightarrow gg} = \frac{\alpha_s^2\alpha_v^2}{16\pi
M^8}\,(N_c^2-1)\chi_s^2C_{S}^2m_{0^+}^3({\bf F_{0^{++}}^S})^2.
\ee
Here and below we make explicit the $SU(3)$-color origin of a factor of $8=N_c^2-1$.

The branching ratios for the decays to the photons, $Z$ and $W^{\pm}$ are
\be
\label{bf s2ff}
\frac{\Gamma_{0^{+}\to \gamma\gamma}}{\Gamma_{0^{+}\to
gg}}=\frac{1}{2}\,\frac{\alpha^2}{\alpha_s^2}\frac{\chi_\gamma^2}{\chi_s^2},
\ee
\be
\frac{\Gamma_{0^{+}\to ZZ}}{\Gamma_{0^{+}\to
gg}}=\frac{1}{2}\,\frac{\alpha_w^2}{\alpha_s^2}\frac{\chi_Z^2}{\chi_s^2}\,\left(1-4\frac{m_Z^2}{m_{0^+}^2}\right)^{1/2}\left(1 - 4 \frac{m_Z^2}{m_{0+}^2}+6\frac{m_Z^4}{m_{0^+}^4}\right),
\ee
\be
\label{bf s2fZ}
\frac{\Gamma_{0^{+}\to \gamma Z}}{\Gamma_{0^{+}\to
gg}}=\frac{1}{4}\,\frac{\alpha\alpha_w}{\alpha_s^2}\frac{\chi_{\gamma
Z}^2}{\chi_s^2}\,\left(1-\frac{m_Z^2}{m_{0^+}^2}\right)^3,
\ee
\be
\label{bf s2WW}
\frac{\Gamma_{0^{+}\to W^+W^-}}{\Gamma_{0^{+}\to
gg}}=\frac{1}{4}\,\frac{\alpha_w^2}{\alpha_s^2}\frac{\chi_W^2}{\chi_s^2}\,\left(1-4\frac{m_W^2}{m_{0^+}^2}\right)^{1/2} \left(1 - 4 \frac{m_W^2}{m_{0^+}^2}+6\frac{m_W^4}{m_{0^+}^4}\right),
\ee
The coefficients $\chi$ used here were defined in Eq.~(\ref{chis}). Factors of $1/2$ in the above ratios come from the color factor
$N_c^2-1=8$ and a difference in the normalization of abelian and
non-abelian generators. An extra $1/2$ is required if the particles in the final state are not identical, such as $W^+W^-$ and $\gamma Z$.

Of course these are SM-tree-level results.  There will be substantial
order-$\alpha_s$ corrections to the $gg$ final state, so the actual
lifetimes will be slightly shorter and the branching fractions to
other final states slightly smaller than given in these formulas.

\paragraph{Decays of the $0^{-+}$ state.} The decay of the pseudoscalar state
$0^{-+}$ to two gauge bosons proceeds in a similar fashion. This decay
is induced by the operator $P$:
\be
\frac{\alpha_i\alpha_v}{M^4}\, \chi_i C_{P}\langle
G^a,G^b|\,\tr{G}_{\mu\nu}{\tilde{G}}^{\mu\nu}|\,0\rangle\,\langle 0|\, P|
0^{-+}\rangle. 
\ee
The amplitude leads to the following two-gluon decay rate:
\be
\label{p2gg}
\Gamma_{0^{-}\rightarrow gg} = \frac{\alpha_s^2\alpha_v^2}{16\pi
M^8}\,(N_c^2-1)\chi_s^2 C_{P}^2m_{0^-}^3({\bf F_{0^{-+}}^P})^2,
\ee
and the same branching fractions as for $0^{++}$, except for the decays
to $ZZ$ and $W^+W^-$,
\be
\frac{\Gamma_{0^{-}\to ZZ}}{\Gamma_{0^{-}\to
gg}}=\frac{1}{2}\,\frac{\alpha_w^2}{\alpha_s^2}\frac{\chi_Z^2}{\chi_s^2}\,\left(1-4\frac{m_Z^2}{m_{0^-}^2}\right)^{3/2},
\ee
\be
\label{bf s2WW2}
\frac{\Gamma_{0^{-}\to W^+W^-}}{\Gamma_{0^{-}\to
gg}}=\frac{1}{4}\,\frac{\alpha_w^2}{\alpha_s^2}\frac{\chi_W^2}{\chi_s^2}\,\left(1-4\frac{m_W^2}{m_{0^-}^2}\right)^{3/2}.
\ee

\begin{figure}[htb]
\begin{center}

\epsfxsize=3.5in \epsffile{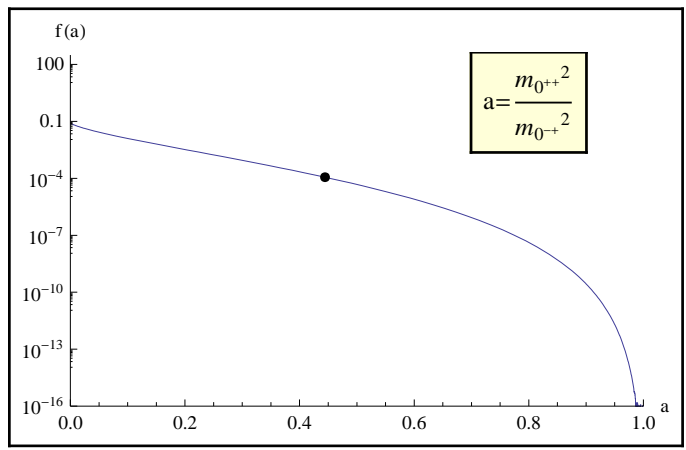}

\end{center}
\vspace{-0.6cm}
\caption{\small Kinematic suppression factor $f(a)$. Point corresponds to a value of $a$ taken for v-glueball masses
from the Morningstar and Peardon spectrum \cite{Morningstar}.}
\label{fig f1}
\end{figure}

The $0^{-+}$ state can also decay to lower lying states by emitting a
pair of gauge bosons, but these decays are
suppressed. For instance, the amplitude for the decay of $0^{-+}\to 0^{++}gg$
is
\be
\frac{\alpha_i\alpha_v}{M^4}\, \chi_i C_{P}\langle
G^a,G^b\,|\,\tr{G}_{\mu\nu}{\tilde{G}}^{\mu\nu}|\,0\rangle\,\langle 0^{++}|\, P|
0^{-+}\rangle \ .
\ee
The matrix element ${\bf M_{0^+0^{-}}^P}=\langle 0^{++}| P|
0^{-+}\rangle$ is a function of the momentum transferred.  Let us first treat
it as approximately constant.  Then we obtain
the decay rate
\be
\label{p2sgg}
\Gamma_{0^{-}\rightarrow 0^{+}+gg} =
\frac{\alpha_s^2\alpha_v^2}{256\pi^3 M^8}\,(N_c^2-1)\chi_s^2C_{P}^2
m_{0^-}^5f(a)({\bf M_{0^+0^{-}}^P})^2,
\ee
where $f$ is
the dimensionless function of the parameter $a=m_{0^+}^2/m_{0^-}^2$,
\be
f(a)=\frac{1}{12}(1-a^2)(1+28a+a^2)+a(1+3a+a^2)\ln a,
\ee
We plot $f$ in figure \ref{fig f1}; it falls rapidly from $1/12$ to
$0$, because of the rapid fall of phase space as the two masses
approach each other.  For the masses in figure~\ref{fig spectrum}, $a=0.44$ and $f\approx
10^{-4}$.  This is in addition to the usual $1/16\pi^2$ suppression of
three-body decays compared to two-body decays.  Thus the branching
fraction for this decay is too small to be experimentally relevant, and our
approximation that the matrix element is constant is inconsequential.
This will be our general conclusion for three-body decays of the
light v-glueball states, and in most cases we will not bother to present
results for such channels.

\paragraph{Decays of the $2^{++}$ state.} Decays of the $2^{++}$ 
glueball to two gauge bosons are induced by more than one operator in
(\ref{eff_lag full2}). In particular, the $2^{++}$ decays due to
the $T_{\mu\nu}$ and $L_{\mu\nu\alpha\beta}$ operators. This
corresponds to the amplitude
\begin{multline}
\frac{\alpha_i\alpha_v}{M^4}\,\chi_i\left[ C_{T}\langle G^a,G^b|\,\tr
{G}_{\mu\alpha}{G}_{\nu}^{~\alpha}|\,0\rangle\,\langle 0|\, T^{\mu\nu}|
2^{++}\rangle  + \right.
\\ \left. + C_{L}\langle
G^a,G^b|\,\tr{G}_{\mu\nu}{G}_{\alpha\beta}|\,0\rangle  \langle 0|\,
L^{\mu\nu\alpha\beta}| 2^{++}\rangle\right].
\end{multline}

The width of the decay to two gluons is
\be
\Gamma_{2^{+}\rightarrow gg} = \frac{\alpha_s^2\alpha_v^2(N_c^2-1)}{160\pi
M^8}\,\chi_s^2m_{2^+}^3\left(\frac{1}{2}\,C_{T}^2 ({\bf
F_{2^{++}}^T})^2 +\frac{4}{3}\,C_{L}^2 ({\bf
F_{2^{++}}^L})^2\right) .
\ee
Here we used the following expressions for the matrix elements:
\be
\langle 0| T^{\mu\nu}| 2^{++}\rangle = {\bf
F_{2^{++}}^T}\,\epsilon^{\mu\nu},
\ee
\be
\label{TMatEl}
\langle 0| L^{\mu\nu\alpha\beta}| 2^{++}\rangle = {\bf F_{2^{++}}^L}
\left[{\mathcal{P}}_{\mu\alpha}\epsilon_{\nu\beta}-{\mathcal{P}}_{\mu\beta}
\epsilon_{\nu\alpha}
+{\mathcal{P}}_{\nu\beta}\epsilon_{\mu\alpha}-{\mathcal{P}}_{\nu\alpha}
\epsilon_{\mu\beta} \right],
\ee
where ${\mathcal{P}}_{\alpha\beta}$ is defined in the caption to table \ref{constant_decay}.

The branching fraction for the decay to two photons is again similar to~(\ref{bf s2ff}). For two Z bosons in the final state, the width of the
decay is equal to
\begin{multline}
\label{t2ZZ}
\Gamma_{2^{+}\to ZZ}=\frac{\alpha_w^2\alpha_v^2}{40\pi
M^8}\,\chi_Z^2{m_{2^+}^3}(1-4\zeta_2)^{1/2}\left(\frac{1}{2}\,C_{T}^2
f_T(\zeta_2)({\bf F_{2^{++}}^T})^2+ \right.
\\ \left. \frac{4}{3}\,C_{L}^2 f_{L}(\zeta_2)({\bf
F_{2^{++}}^L})^2+
\frac{40}{3}\,C_{T}C_{L}f_{TL}(\zeta_2){\bf
F_{2^{++}}^T}{\bf F_{2^{++}}^L}\right),
\end{multline}
where $f_T$, $f_L$, $f_{TL}$ are the following functions of the parameter
$\zeta_2=m_Z^2/m_{2^+}^2$.
\be
f_T(\zeta_2)= 1-3\zeta_2 + 6\zeta_2^2\ , \ f_L(\zeta_2)=1+2\zeta_2 +
36\zeta_2^2\ , \ f_{TL}(\zeta_2)=\zeta_2(1-\zeta_2)\ .
\ee

The decay to $W^+W^-$ is obtained from Eq.~(\ref{t2ZZ}) by substituting
$\chi_Z\to \chi_W$, $m_Z\to m_W$ and multiplying by $1/2$. For the
$\gamma Z$ final state, the decay rate is
\begin{multline}
\label{t2Zgamma}
\Gamma_{2^{+}\to \gamma Z}=\frac{\alpha\alpha_w\alpha_v^2}{80\pi
M^8}\,\chi_{\gamma Z}^2{m_{2^+}^3}(1-\zeta_2)^3\left(\frac{1}{2}\,C_{T}^2
g_T(\zeta_2)({\bf F_{2^{++}}^T})^2+ \right.
\\ \left. \frac{4}{3}\,C_{L}^2 g_{L}(\zeta_2)({\bf
F_{2^{++}}^L})^2+  \frac{20}{3}\,C_{T}C_{L}\zeta_2{\bf
F_{2^{++}}^T}{\bf F_{2^{++}}^L}\right),
\end{multline}
where
\be
g_T(\zeta_2)= 1+\frac12\,\zeta_2 + \frac16\,\zeta_2^2, \qquad
g_L(\zeta_2)=1+3\zeta_2 + 6\zeta_2^2.
\ee

As in the case of the $0^{-+}$, we can ignore the three-body
transitions $2^{++}\to 0^{++}+gg$, etc.

\paragraph{Decays of the $2^{-+}$ state.} The dominant decays of the $2^{-+}$
state occur due to the $L_{\mu\nu\alpha\beta}$ operator. The amplitude for
such decays is given by
\begin{equation}
\frac{\alpha_i\alpha_v}{M^4}\,\chi_i C_{L}\langle G^a,G^b|\,\tr
{G}_{\mu\nu}{G}_{\alpha\beta}|\,0\rangle  \langle 0|\,
L^{\mu\nu\alpha\beta}|\, 2^{-+}\rangle.
\end{equation}
The correct Lorentz structure that singles out the negative parity part
of the operator $L_{\mu\nu\alpha\beta}$ is as follows:
\begin{multline}
\langle 0| L^{\mu\nu\alpha\beta}|\, 2^{-+}\rangle = {\bf
F_{2^{-+}}^L}\left(\epsilon_{\mu\nu\rho\sigma}\epsilon^\sigma_{~\beta}n^\rho
n_\alpha - \epsilon_{\mu\nu\rho\sigma}\epsilon^\sigma_{~\alpha}n^\rho
n_\beta +\right.
\\ \left.
+ \epsilon_{\alpha\beta\rho\sigma}\epsilon^\sigma_{~\nu}n^\rho n_\mu -
\epsilon_{\alpha\beta\rho\sigma}\epsilon^\sigma_{~\mu}n^\rho n_\nu\right),
\end{multline}
where $n_\mu=p^{\,\mu}/m_{2^-}$ is a unit vector in the direction of the 4-momentum of the v-glueball.

The decay rate to two gluons
is then given by
\be
\label{l2gg}
\Gamma_{2^{-}\to gg} = \frac{\alpha_s^2\alpha_v^2}{120\pi
M^8}\,(N_c^2-1){\chi_s}^2m_{2^-}^3C_{L}^2  ({\bf
F_{2^{-+}}^L})^2
\ee
and $\Gamma_{2^{-}\to\gamma\gamma}$ is provided by the same relation as
(\ref{bf s2ff}). The widths of the decay to $ZZ$ and $\gamma Z$ can be
found from the ratios
\be
\label{l2ggZ}
\frac{\Gamma_{2^{-}\to ZZ}}{\Gamma_{2^{-}\to
gg}}=\frac{1}{2}\,\frac{\alpha_w^2}{\alpha_s^2}\frac{\chi_Z^2}{\chi_s^2}\,\left(1-4\frac{m_Z^2}{m_{2^-}^2}\right)^{1/2}
\left(1+2\frac{m_Z^2}{m_{2^-}^2} - 24 \frac{m_Z^4}{m_{2^-}^4}\right)\ ,
\ee
\be
\frac{\Gamma_{2^{-}\to \gamma Z}}{\Gamma_{2^{-}\to
gg}}=\frac{1}{4}\,\frac{\alpha\alpha_w}{\alpha_s^2}\frac{\chi_{\gamma
Z}^2}{\chi_s^2}\,\left(1-\frac{m_Z^2}{m_{2^-}^2}\right)^3\left(1+3\frac{m_Z^2}{m_{2^-}^2}
+ 6 \frac{m_Z^4}{m_{2^-}^4}\right)\ ,
\ee
and the width for the decay to $W^+W^-$ is again obtained by
substituting in (\ref{l2ggZ}) $\chi_Z\to\chi_W$, $m_Z\to m_W$ and
dividing the result by 2.

As before, we can neglect 3-body decays.  However, there is a 2-body
radiative decay that we should consider, although, as we will see,
for the masses in figure~\ref{fig spectrum} it is of the same order as the
3-body decays.  For the $SU(3)$ spectrum in
\cite{Morningstar} (and possibly all pure glue $SU(n)$, $n\geq 3$)
the $2^{-+}$ state is slightly heavier than the lightest state in the $C$-odd
sector, the pseudovector $1^{+-}$. Thus, we need at least to consider
the decay $2^{-+}\to 1^{+-}+\gamma$. This decay is induced by the
second type of operators (table \ref{dim6}) in the effective action
(\ref{eff_lag full2}). The amplitude of the decay reads \be
\label{2-+amp0}
\frac{eg_v^3}{(4\pi)^2M^4}\, \chi\langle
\gamma|\,G^{\mu\nu}|\,0\rangle \, \left(C_{\Omega^{(1)}} \langle
1^{+-}|\,\Omega_{\mu\nu}^{(1)}|\,2^{-+}\rangle + C_{\Omega^{(2)}} \langle
1^{+-}|\,\Omega_{\mu\nu}^{(2)}|\,2^{-+}\rangle\right)\ .
\ee

Unfortunately nothing quantitative is known about matrix elements like
$\langle 1^{+-}|\Omega_{\mu\nu}^{(n)}|\,2^{-+}\rangle$.  In fact each
contains multiple Lorentz structures, constructed out of polarization
tensors $\epsilon_{\alpha}$, $\epsilon_{\beta\gamma}$ and momenta $p$
and $q$ of the $1^{+-}$ and $2^{-+}$ v-glueballs, times functions of
the momentum transfer, cf.~\cite{Delbourgo:1994ab}. Some simplification can be made if one takes
into account the fact that masses of the v-glueballs are close, which
we will assume below.

We start by writing the general expression for the
amplitude~(\ref{2-+amp0}):
\begin{multline}
\label{2-+amp}
\langle\, \gamma|\,G^{\mu\nu}|\,0\rangle \, \langle\,
1^{+-}|\,\Omega_{\mu\nu}^{(n)}|\,2^{-+}\rangle = 2{\bf
M_{2^{-+}1^{+-}}^{\Omega^{(n)}}}(k\cdot
p\,\varepsilon^\alpha\epsilon_{\alpha\beta}\epsilon^\beta -
p\cdot\varepsilon\,k^\alpha\epsilon_{\alpha\beta}\epsilon^{\beta}) +
\\  +
2{\bf M_{2^{-+}1^{+-}}^{\Omega^{(n)}\prime}}(k\cdot
p\,\varepsilon\cdot\epsilon - k\cdot \epsilon
p\cdot\varepsilon)\,\frac{p^\alpha\epsilon_{\alpha\beta}p^\beta}{m_{2^{-}}^2} +
\\ + 2{\bf M_{2^{-+}1^{+-}}^{\Omega^{(n)}\prime\prime}}(k\cdot
p\,\varepsilon^\alpha\epsilon_{\alpha\beta}p^\beta - p\cdot \varepsilon
\,k^\alpha\epsilon_{\alpha\beta}p^\beta)\frac{q\cdot\epsilon}{m_{2^{-}}^2}\,,
\end{multline}
where $n=1,2$ and $k$, $\varepsilon_\alpha$ are the momentum and
polarization of the photon. All contributions of the terms proportional to primed
form-factors (which correspond to higher partial waves) are suppressed by powers of
$(m_{2^{-}}-m_{1^{+}})/(m_{2^{-}}+m_{1^{+}})\simeq 0.017$, so we may
neglect them.\footnote{Here we assume that the primed form-factors
$\mathbf{M}$ are at most of the same order of magnitude as ${\bf
M_{2^{-+}1^{+-}}^{\Omega^{(n)}}}$.} Note, however, that if the mass splitting is much larger for $n_v>3$, then there will be
additional unknown quantities that will modify our result below.

We now find
\begin{multline}\label{2m1pgam}
\Gamma_{2^{-}\to 1^{+}+\gamma}=\frac{\alpha\alpha_v^3}{960\pi
M^8}\,{\chi^2}\,\frac{(m_{2^-}^2-m_{1^+}^2)^3}{m_{2^-}^5m_{1^+}^2}\,\times
\\ \times(3m_{2^-}^4 + 34m_{2^-}^2m_{1^+}^2 + 3
m_{1^+}^4)\left({\bf M_{2^{-+}1^{+-}}^{\Omega}}\right)^2.
\end{multline}
Here we introduced the notation
\be
\label{momega}
{\bf M_{2^{-+}1^{+-}}^{\Omega}} \equiv C_{\Omega^{(1)}}{\bf M_{2^{-+}1^{+-}}^{\Omega^{(1)}}} +
C_{\Omega^{(2)}}{\bf M_{2^{-+}1^{+-}}^{\Omega^{(2)}}}.
\ee
Since the the form-factors ${\bf
M_{2^{-+}1^{+-}}^{\Omega^{(n)}}}$ are unknown, we shall not distinguish between
them and will use a collective notation, similar to (\ref{momega}), for
them in the future. In the same manner we will use the notation
\be
\Omega_{\mu\nu}\equiv C_{\Omega^{(1)}}\Omega^{(1)}_{\mu\nu} +
C_{\Omega^{(2)}}\Omega^{(2)}_{\mu\nu}.
\ee

The factor $(m_{2^-}^2-m_{1^+}^2)^3$ strongly suppresses the
amplitude, given the spectrum of figure~\ref{fig spectrum}, and a
rough estimate suggests it is of the same size as the three-body
decays of the $2^-$ state, and consequently negligible.  However this
splitting is so small that it is sensitive to numerical uncertainties
in the lattice calculation, and might well be different for other
gauge groups.  In particular, this decay channel might be closed, or
might be more widely open than suggested by figure~\ref{fig spectrum},
depending on the mass spectrum.

Given the uncertainty on the spectrum and the unknown v-glueball
mass scale, it is worth noting that the radiative decay of
$2^{-+}$ to $1^{+-}$ can in principle occur through an emission of the
$Z$ boson.  This decay is
slightly more involved than the decay with photon emission considered
above. Additional unknown form factors related to the finiteness of
the $Z$ mass further reduce the predictive power of any
computation. But such a decay may be forbidden by kinematics, and if
allowed it is probably of little importance for the discovery of
v-glueballs.  Its rate will almost certainly lie somewhere between 0
and $\tan^2\theta_W\sim 20$\% of the rate for decays to a
photon. There is no reason for the form factors ${\bf M}(k^2)$ to be
enhanced at $k^2\simeq m_Z^2$.  Since the $Z$ boson has only a few
percent branching fraction to electrons and muons, the ratio of identifiable
$Z$ decays to photon decays is less than 2\%.  We therefore will not
present formulas for this decay mode.

Again we emphasize that in obtaining the results (\ref{2m1pgam}) we
made some assumptions and approximations, including $\Delta m\ll
m_{2^-}$, and these results may require generalization in other
calculations.  However, we will adhere to similar simplifying
approximations in the other radiative decays computed below.

\subsection{Decays of the vector and pseudovector}
\label{vector glueballs}
In the $C$-odd sector, the lightest
v-glueballs are the pseudovector $1^{+-}$ and vector $1^{--}$.  The
lowest-dimension operators that can create or destroy $1^{--}$ and
$1^{+-}$ v-glueballs are the $d=6$ $\Omega_{\mu\nu}$ operators (table
\ref{dim6}).  Direct annihilation to non-abelian SM gauge bosons would
require an operator in the effective action of dimension $D=12$, and
is hence negligible.  Instead these operators, combined with a
hypercharge field strength tensor to form an operator of dimension 8,
induce radiative decays to $C$-even v-glueballs, and potentially, for
the $1^{--}$ state, annihilation to SM fermions via an off-shell
$\gamma$ or $Z$.  Three-body decays induced by the dimension 4
operators $S,P,T,L$, although quite uncertain
because of the presence of many decay channels with many form factors,
appear to be sufficiently suppressed by phase
space that they can be disregarded.

Below, we will generally not write formulas for radiative decays by $Z$
emission.  As we discussed for the $2^{-}\to 1^{+}+\gamma/Z$ decay,
the ratio of leptonic $Z$ bosons to photons is unlikely to reach
$2\%$, even if there is no phase space suppression (which there
typically is.)  Moreover, decays to $Z$ are described by a larger
number of unknown form factors, making any attempt to predict the
corresponding
decay widths and branching ratios even more uncertain than for photon
emission.

\paragraph{Decays of the $1^{+-}$ state.} Since the $1^{+-}$ is the lightest
v-glueball in the $C$-odd sector, it can only decay, radiatively, to the
lighter v-glueballs in the $C$-even sector.

According to table~\ref{transition_matrix}, the amplitude of the decay $1^{+-}\rightarrow 0^{++}+ \gamma$ is given by\footnote{Similar amplitudes are used in the studies of vector and pseudovector mesons. See for example~\cite{Gao}, \cite{Rapp} and~\cite{Kaiser}.}
\be
\frac{eg_v^3}{(4\pi)^2M^4}\,{\chi}\langle\, \gamma|\, G^{\mu\nu}|\, 0
\rangle \, \langle 0^{++}|\, \Omega_{\mu\nu}|\,1^{+-}\rangle =
\frac{eg_v^3\chi}{(4\pi)^2M^4}\,
2k_{\mu}\varepsilon_{\nu} \epsilon^{\mu\nu\alpha\beta}p_\alpha  \epsilon_\beta
{\bf M_{1^{+-}0^{++}}^\Omega},
\ee
where $\varepsilon_\mu $ and $\epsilon_\mu$ are the polarization
vectors of the photon and the pseudovector v-glueball respectively;
$p_\mu$ is the 4-momentum of the $0^{++}$. The Levi-Civita tensor assures
the final particles are in a p-wave, as required by parity
conservation. The decay rate of this process is
\be
\label{h2sg}
\Gamma_{1^{+}\to 0^{+}+\gamma}= \frac{\alpha\alpha_{v}^3}{24\pi
M^8}\,{\chi^2}\,\frac{(m_{1^+}^2-m_{0^+}^2)^3}{
m_{1^+}^3}\,({\bf M_{1^{+-}0^{++}}^\Omega})^2.
\ee

In the case of the decay to the pseudoscalar v-glueball $1^{+-}\rightarrow 0^{-+}+ \gamma$, the amplitude is given by
\begin{multline}
\label{1+-2pg}
\frac{e g_v^3}{(4\pi)^2M^4}\,\chi\langle \,\gamma|\, G^{\mu\nu}|\, 0
\rangle \, \langle 0^{-+}|\, \Omega_{\mu\nu}|\,1^{+-}\rangle =
\\ =
\frac{eg_v^3}{(4\pi)^2M^4}\,\chi \ 2 k^{\mu}\varepsilon^{\nu}(p_{\mu} \epsilon_\nu
- p_\nu \epsilon_\mu){\bf M_{1^{+-}0^{-+}}^\Omega},
\end{multline}
where $p_\mu$ is the 4-momentum of the $0^{-+}$. The rate of the decay to the pseudoscalar is then
\be
\label{h2pg}
\Gamma_{1^{+}\to 0^{-} + \gamma} = \frac{\alpha\alpha_{v}^3}{24\pi
M^8}\,{\chi^2}\,\frac{(m_{1^+}^2-m_{0^-}^2)^3}{
m_{1^+}^3}\,({\bf M_{1^{+-}0^{-+}}^\Omega})^2.
\ee

The ratio of the decay rates to $0^{-+}$ and $0^{++}$ is
\be
\frac{\Gamma_{1^{+}\rightarrow
0^{-}+\gamma}}{\Gamma_{1^{+}\rightarrow 0^{+}+\gamma}} =
\left(\frac{m_{1^+}^2-m_{0^-}^2}{m_{1^+}^2-m_{0^+}^2}\right)^3\left(\frac{{\bf
M_{1^{+-}0^{-+}}^\Omega}}{{\bf M_{1^{+-}0^{++}}^\Omega}}\right)^2.
\ee
For the spectrum of figure~\ref{fig spectrum}, the factor involving
the masses is about 0.39; the ratio of matrix elements is unknown, but
if we guess that ${{\bf M_{1^{--}0^{\pm+}}^\Omega}}\sim 1/{\bf
F^{S,P}_{0^{\pm+}}}$, as would be true for pion emission, and use the lattice
results from \cite{Morningstar2}, we would
find this ratio to be slightly larger than 1.  In any case, there is
no sign of a significant suppression of one rate relative to the
other.

Finally, in the case of the decay to the tensor v-glueball, the amplitude $1^{+-}\rightarrow 2^{++}+ \gamma$  contains two independent form factors in the lowest partial wave approximation, denoted ${\bf M_{1^{+-}0^{-+}}^\Omega}$ and ${\bf {M'}_{1^{+-}0^{-+}}^\Omega}$,
\begin{multline}
\label{1+-2tg}
\frac{e g_v^3}{(4\pi)^2M^4}\,\chi\langle \,\gamma|\, G^{\mu\nu}|\, 0
\rangle \, \langle 2^{++}|\, \Omega_{\mu\nu}|\,1^{+-}\rangle =
\\ =
\frac{e g_v^3}{(4\pi)^2M^4}\,\chi \ 2 k^{\mu}\varepsilon^{\nu} \epsilon_{\mu\nu\alpha\beta} \epsilon_{\beta\lambda}( \epsilon_{\lambda}p_{\alpha} {\bf M_{1^{+-}0^{-+}}^\Omega}+ \epsilon_{\alpha}p_{\lambda} {\bf {M'}_{1^{+-}0^{-+}}^\Omega})
\end{multline}
and the corresponding decay rate is
\begin{multline}
\label{1+-2tgph}
\Gamma_{1^{+}\to 2^{+}+\gamma}= \frac{\alpha\alpha_{v}^3}{576\pi
M^8}\,{\chi^2}\,\frac{(m_{1^+}^2-m_{2^+}^2)^3}{
m_{1^+}^5 m_{2^+}^2}\,
 \left(3 m_{2^+}^4+34 m_{1^+}^2 m_{2^+}^2+3 m_{1^+}^4\right) \times  \\ \times \left[\left({\bf M_{1^{+-}2^{++}}^\Omega}+ {\bf {M'}_{1^{+-}2^{++}}^\Omega} f(m_{1^+},m_{2^+}) \right)^2 \right.
\\ \left.+ \left({\bf {M'}_{1^{+-}2^{++}}^\Omega}\right)^2 g(m_{1^+},m_{2^+}) \right],
\end{multline}
where $f$ and $g$ are the following functions of the v-glueball masses,
\be
\label{fmm}
f(m_{1^+},m_{2^+})=\frac{\left(m_{2^+}^2-m_{1^+}^2\right)\left(3 m_{2^+}^2+7 m_{1^+}^2\right)}{3 m_{2^+}^4+34 m_{1^+}^2 m_{2^+}^2+3 m_{1^+}^4},
\ee
\be
\label{gmm}
g(m_{1^+},m_{2^+})= 12\,\frac{\left(m_{2^+}^2-m_{1^+}^2\right)^2 m_{1^+}^2(6 m_{2^+}^4+8 m_{1^+}^2 m_{2^+}^2+m_{1^+}^4) }{m_{2^+}^2 \left(3 m_{2^+}^4+34 m_{1^+}^2 m_{2^+}^2+3 m_{1^+}^4 \right)^2}.
\ee

\paragraph{Decays of the $1^{--}$ state.} The decays of the vector
v-glueball are similar to the decays of the pseudovector, with a few
additions. In contrast to the case of the $1^{+-}$, the $1^{--}$
can annihilate through an off-shell vector boson
to a SM fermion-antifermion pair.
But the radiative decays to light v-glueballs in the
$C$-even sector still typically dominate.

The radiative decay to the scalar,
$1^{--}\to 0^{++}+ \gamma$, is analogous to the decay $1^{--}\to
0^{++}+\gamma$; see table~\ref{transition_matrix}
and~(\ref{h2pg}). Thus, its rate is
\be
\label{1--tog}
\Gamma_{1^{-}\to 0^{+} + \gamma} = \frac{\alpha\alpha_{v}^3}{24\pi
M^8}\,{\chi^2}\,\frac{(m_{1^-}^2-m_{0^+}^2)^3}{
m_{1^-}^3}\,({\bf M_{1^{--}0^{++}}^\Omega})^2.
\ee

The decay to the pseudoscalar
is analogous to the decay~(\ref{h2sg}) and has the rate
\be
\label{rho2pg}
\Gamma_{1^{-}\to 0^{-}+\gamma}= \frac{\alpha\alpha_{v}^3}{24\pi
M^8}\,{\chi^2}\,\frac{(m_{1^-}^2-m_{0^-}^2)^3}{
m_{1^-}^3}\,({\bf M_{1^{--}0^{-+}}^\Omega})^2.
\ee

The amplitude of the decay to the $2^{++}$ state is similar to the
amplitude~(\ref{2-+amp}) of the decay $2^{-+}\to 1^{+-}+\gamma$.
However, in this case the masses of the two states are not close, and
our approximation which allowed us to ignore the contribution of three
additional form factors is not valid.  We therefore restrict ourselves
to just demonstrating the general expression for the amplitude.
\begin{multline}
\label{1--2tg}
\frac{e g_v^3\chi}{(4\pi)^2M^4}\langle \,\gamma|\, G^{\mu\nu}|\, 0
\rangle  \langle 2^{++}|\, \Omega_{\mu\nu}|\,1^{--}\rangle  = 2{\bf
M_{1^{--}2^{++}}^{\Omega}}(k\cdot
p\,\varepsilon^\alpha\epsilon_{\alpha\beta}\epsilon^\beta -
p\cdot\varepsilon\,k^\alpha\epsilon_{\alpha\beta}\epsilon^{\beta})
\\ +
2{\bf M_{1^{--}2^{++}}^{\Omega\prime}}(k\cdot
q\,\varepsilon\cdot\epsilon - k\cdot \epsilon
q\cdot\varepsilon)\,\frac{q^\alpha\epsilon_{\alpha\beta}q^\beta}{m_{1^{-}}^2} +
\\ + 2{\bf M_{1^{--}2^{++}}^{\Omega\prime\prime}}(k\cdot
q\,\varepsilon^\alpha\epsilon_{\alpha\beta}q^\beta - q\cdot \varepsilon
\,k^\alpha\epsilon_{\alpha\beta}q^\beta)\frac{p\cdot\epsilon}{m_{1^{-}}^2}.
\end{multline}
A complete formula for the decay rate is not very useful, given the
large number of unknown form factors that enter.

The $1^{--}$ state is also massive enough to decay to the $2^{-+}$
state. This decay has an amplitude similar to the decay $1^{+-}\to
2^{++}+\gamma$, given in (\ref{1+-2tg}).  One can find the decay rate
\begin{multline}
\label{1--2tgph}
\Gamma_{1^{-}\to 2^{-}+\gamma}= \frac{\alpha\alpha_{v}^3}{576\pi
M^8}\,{\chi^2}\,\frac{(m_{1^-}^2-m_{2^-}^2)^3}{
m_{1^-}^5 m_{2^-}^2}\,
 \left(3 m_{2^-}^4+34 m_{1^-}^2 m_{2^-}^2+3 m_{1^-}^4\right) \times  \\ \times \left[\left({\bf M_{1^{--}2^{-+}}^\Omega}+ {\bf {M'}_{1^{--}2^{-+}}^\Omega} f(m_{1^-},m_{2^-}) \right)^2 \right.
\\ \left.+ \left({\bf {M'}_{1^{--}2^{-+}}^\Omega}\right)^2 g(m_{1^-},m_{2^-}) \right],
\end{multline}
where functions $f$ and $g$ are defined by~(\ref{fmm}) and (\ref{gmm}) respectively.

Now we consider the decay of the $1^{--}$ to SM fermion pairs through
an off-shell $\gamma$ or $Z$.  For large $m_{1^-}$ we can neglect the
$Z$ mass and treat the radiated particle as an off-shell hypercharge
boson. The amplitude reads
\be
\frac{\alpha g^3_v}{2\pi M^4}\,\frac{\chi}{\cos^2\theta_W}\,\langle
f,\bar{f}| Y_L\bar{\psi}_L\gamma^\mu\psi_L +
Y_R\bar{\psi}_R\gamma^\mu\psi_R|\, 0 \rangle \, \frac{1}{p^2}\, \langle
0| p^{\nu}\Omega_{\nu\mu}|1^{--}\rangle.
\ee
Here $Y_L$ and $Y_R$ are left and right hypercharges of the
emitted fermions. The matrix element of $\Omega_{\mu\nu}$ can be read off from 
table~\ref{constant_decay}. The width (ignoring the
fermion masses) is given by
\be
\label{rho2ll}
\Gamma_{1^{-}\to \bar{f}f}= \frac{{2\alpha^2}\alpha_v^3}{3
M^8}\,\frac{\chi^2}{\cos^4\theta_W}\,(Y_L^2+Y_R^2)\,m_{1^{-}}^3({\bf
F_{1^{--}}^\Omega})^2.
\ee
For quarks a factor of 3 must be included to account for color.

The above result is valid for $m_{1^-}\gg m_Z$.
For smaller $m_{1^-}$ one must account for the
non-zero $Z$ mass through the substitution
\begin{multline}
\label{Zpoleeffect}
\frac{(Y_L^2+Y_R^2)}{\cos^4\theta_W}\to \left(Q -
\frac{Q\cos^2\theta_W- Y_L}{\cos^2\theta_W}\,\frac{m_{1^-}^2}{m_{1^-}^2-m_Z^2}\right)^2+
\\ + Y_R^2\left(1 +
\frac{\sin^2\theta_W}{\cos^2\theta_W}\,\frac{m_{1^-}^2}{m_{1^-}^2-m_Z^2}\right)^2,
\end{multline}
which accounts for a finite mass of the $Z$-boson. Here $Q=T_3+Y$ is the charge of $f$.
A quick check shows
that this rate, whose ratio to
radiative decays
is (for large $m_{1^-}$)
\be
\frac{\Gamma_{1^{-}\to\gamma^*/Z^*\to
f\bar{f}}}{\Gamma_{1^{-}\rightarrow 0^{+}+\gamma}} = \frac{16 \pi \alpha}{\cos^4\theta_W}\,(Y_L^2+Y_R^2)
\left(\frac{m_{1^-}^2}{m_{1^-}^2-m_{0^+}^2}\right)^3\left(\frac{{\bf
F_{1^{--}}^\Omega}}{{\bf M_{1^{--}0^{++}}^\Omega}}\right)^2 ,
\ee
is not negligible.  The first factor in curved brackets is a factor of
a few, while the second factor in curved brackets may be large,
especially at large $n_v$.  Decays to electrons and muons
will be reconstructable as a resonance, so despite the uncertain branching
fractions
this decay mode is worthy of careful consideration.

Decay of the $1^{--}$ state to the $1^{+-}$ v-glueball can only
proceed with the emission of at least two SM gauge bosons. Although
such decays are suppressed, the details of the calculation for 
$1^{--}\to 1^{+-}+gg$ are presented in the Appendix.

\subsection{Decays of the remaining states}

We may infer without detailed calculation that
the likely decays of the other v-glueballs in the $C$-odd sector are
radiative.  Three-body decays to two gauge bosons plus another $C$-odd
v-glueball are quite suppressed by phase space, because the mass
splittings in the $C$-odd sector are never large.  Even the splitting
of the $0^{+-}$ state from the $1^{+-}$ state is only 1.1 $m_{0^{++}}$.
By contrast, two-body radiative decays into the $C$-even sector have
significantly larger phase space.  (In the appendix, we confirm this
for decays of the $1^{--}$ state.)  Meanwhile, no operator appearing
in the effective action at dimension $D=8$ permits the $0^{+-}$,
$2^{\pm -}$, or $3^{\pm -}$
states to annihilate directly to standard model particles.  Therefore,
we should expect that {\it all} of these states decay radiatively,
emitting typically a photon or more rarely (if kinematically allowed)
a $Z$, to a v-glueball of opposite $C$.  Their lifetimes will be of order
or slightly shorter than that of the $1^{-+}$, due to enhanced phase
space and additional decay channels.

The $3^{++}$ state is more complicated.  No operator allows it to
annihilate directly to standard model gauge bosons, so it will decay
either by a two-body radiative transition to the $C$-odd sector or by
a three-body decay to two gauge bosons plus a $C$-even v-glueball.
For this state, in contrast to the $C$-odd states, the mass splittings
tend to suppress the radiative decay and enhance the three-body
decays.  With many contributing decay channels and unknown form
factors, it seems impossible to estimate which type of decay is
dominant.  Indeed simple estimates suggest they are of the same order,
with large uncertainties.  Qualitatively, if the colored $X$
particles are very heavy and $\chi_s$ is very small, radiative decays
will probably dominate, while tight degeneracies within the $X$ multiplet(s)
could suppress $\chi$ and reverse the situation. But quantitative
prediction seems impossible.


\section{Conclusions}
\label{sec conclusions}

Let us first summarize our results and their immediate implications.
\begin{itemize}
\item We have seen that annihilation decays dominate those states that
can be created by dimension $d=4$ operators (the $0^{\pm +}$ and
$2^{\pm +}$).  Their branching fractions are dominated by decays to
$gg$, with decays to $\gamma\gamma$ having a branching fraction of
$\sim 0.4\%$, assuming the $X$ fields form complete $SU(5)$ multiplets
of equal mass.  If the colored $X$ particles are much heavier than the
uncolored ones, then decays to electroweak bosons can dominate.
\item
Most other states decay by radiatively emitting a photon, or (at
a rate that is at most $\tan^2\theta_W$ compared to photon emission)
a $Z$ boson.
\item
The $1^{--}$ is a special case; it typically prefers to decay
radiatively but has a non-negligible annihilation decay to an
off-shell $\gamma$ or $Z$.
\item
The $3^{++}$ is also special; three-body decays to gluons plus
a $C$-even v-glueball could be of the same order or even dominate
over radiative decays to the $C$-odd sector.
\item
In all, we expect the final states from v-glueball production to be
rich in jets and stray photons, with occasional photon pairs,
leptons and some missing
energy from neutrinos.  The two-photon resonances from the
annihilation decays of $C$-even v-glueballs are likely to be the
discovery signatures, along with the $\gamma\gamma\gamma$ and $\gamma
g g$ resonances from cascade decays of $C$-odd v-glueballs.
\item
Depending on the parameters, the lifetime of any given state can vary
over many orders of magnitude.  But for any fixed choice of
parameters, lifetimes of the v-glueballs vary over at least three or
four orders of magnitude, the details depending on unknown v-glueball
matrix elements and mass ratios, as well as the $X$ mass spectrum.
Displaced vertices can potentially serve as a discovery channel.
\item
There are several opportunities for discovery of this signal in
displaced vertices.  One option arises from $gg$ decays in events
triggered by photons, another from $W^+W^-$ decays triggered by the
muon or electron in a leptonic $W$ decay, and a third from photons
that arrive late or (if converted) point away from the primary
vertex.
\end{itemize}

Our results are robust, but some cautionary and clarifying remarks are
in order.  Clearly, numerical application of our formulas is
currently subject to considerable uncertainties, due especially to the
many unknown matrix elements that arise, and due also to the unknown
spectrum for gauge groups other than $SU(3)$.  Of course these
uncertainties are largely reducible through additional lattice gauge
theory computations, should a discovery of a sector of this type be
made.  However, there are other potential subtleties to keep in mind.
If the $X$ fields and the v-glueball states have comparable masses,
then mixing between these states cannot be neglected.  This could lead
to additional physical effects that we have not considered.  We also
remind the reader that we have worked at leading non-vanishing order
and that higher-order corrections are not negligible when precise predictions
are required.

A more qualitative uncertainty, and an interesting opportunity, arises
from the gauge group.  For $SU(n)$, $n>2$, it is anticipated that the
glueball spectrum is similar to that of $SU(3)$, as calculated by
\cite{Morningstar}.  However the $SU(2)$ spectrum, and more generally that of
any $Sp(2n_v)$ or $SO(2n_v+1)$ gauge group, has no $C$-odd sector.
The operators $\Omega_{\mu\nu}^{(i)}$ do not exist, as they are built
from the $d^{abc}$ symbol absent from such groups, and the
corresponding $C$-odd states are also absent.

For $SO(2n_v)$ the situation is more subtle.  The first cases are
$SO(4)$, which is not simple and has two sets of $SU(2)$ v-glueballs,
and $SO(6)$, which is the same as $SU(4)$.  The $d=6$
$\Omega_{\mu\nu}^{(i)}$ operators are present for $SO(6)$, but for
general $SO(2n_v)$ the $\Omega_{\mu\nu}^{(i)}$ operators become
Pfaffian operators of dimension $2n_v$, built from a single epsilon
symbol and $n_v$ field strengths.  As suggested by~\cite{Jaffe} and as
verified by~\cite{Morningstar}, there is a correlation in the QCD
spectrum and in the glueball spectrum between the dimension of an
operator and the mass of the lightest corresponding state.  For this
reason we expect that for a pure $SO(2n_v)$ gauge theory with $n_v>3$,
the $C$-odd states are heavier than in figure~\ref{fig spectrum}
relative to the $C$-even states.  Their production rate is likely to
be quite suppressed as a result, but are still interesting, since
several are likely to be
unable to decay to other v-glueballs alone, and
will be metastable.  Certainly the lightest $C$-odd state (probably
still the $1^{+-}$) cannot decay to two or more
$C$-even v-glueballs, so it will likely decay by radiating a photon or $Z$.  
Moreover, the degeneracy of the light $C$-odd states seen in
$SU(3)$ may well persist more generally, making these states
potentially unable to decay to two v-glueballs in a $C$-odd final
state, such as $1^{+-} + 0^{++}$.  All of these states will decay therefore
to the $C$-even sector by radiating a photon (or $Z$), except the $1^{--}$ that
may again decay to standard model fermions.  The larger
phase space for $n_v>3$ means the lifetimes may be much shorter than
those of the $C$-even states, a fact which could be phenomenologically
important if
$\Lambda_v$ and $\Lambda_v/M$ are so small 
that the $C$-even states are unobservably long-lived.

Thus study of the spectrum of the v-glueballs may provide some
information on the gauge group.  Combined with some partial
information about the $X$ production rate and the branching
fractions of $X\bar X$ annihilations, it may well be possible to identify
the gauge group precisely.

Finally, we have assumed here that the v-glueballs are the low-energy
degrees of freedom of an asymptotically weakly-coupled gauge theory.
The AdS/CFT correspondence \cite{AdS/CFT,AdS/CFT2} 
allows us to learn what
one might observe if the theory has a large 't Hooft coupling in the
ultraviolet.  In particular, the low-lying glueballs of such a theory
can be described as modes of a string theory
on a 10-dimensional space compactified to 5 dimensions.
Such a theory \cite{Wittenmodel,glueballs1,glueballs2} 
will have light scalars, pseudoscalars, tensors, etc.,
but will not have any light $2^{-+}$ state.  Apparently
the mass of this state may serve as a crude probe of the size of the
ultraviolet 't Hooft coupling, as long as its mass
is not so high as to render
the state unstable to decay to lighter glueballs.

Our formulas now permit a variety of phenomenological studies.  The
issue of LHC searches will be considered in \cite{vglue2}.

\section*{Acknowledgments}

We thank Y.~Gershtein, C.~J.~Morningstar, A.~Yu.~Morozov, M.~E.~Peskin, M.~Shifman and A.~Vainshtein for
conversations. This work was supported in part by the Department of Energy
grant DE-FG02-96ER40949. The work of D.M. was also partly supported by
the RFBR grant 07-02-01161, the grant for Support of Scientific
Schools NSh-3035.2008.2, the center of excellence supported by the
Israel Science Foundation (grant No. 1468/06), the grant DIP H52 of
the German Israel Project Cooperation, the BSF United-States-Israel
binational science foundation grant 2006157 and the German Israel
Foundation (GIF) grant No. 962-94.7/2007.

\

\

\section*{Appendix}
\begin{appendix}

In the main body of the text we have argued that three-body decays  of states
in the $C$-odd sector are largely suppressed as compared to their
radiative decays to light $C$-even states.  As an example, here we consider a three-body decay $1^{--}\to 1^{+-}gg$ and demonstrate that there is a substantial suppression of its rate. We will restrict ourselves to consider the case of the s-wave decay mode since this is
expected to give the highest contribution to the decay rate in a partial wave expansion. In this approximation only $P$ and $L_{\mu\nu\alpha\beta}$ operators contribute (table~\ref{dim4}). This corresponds to the amplitude
\begin{multline*}
\frac{\alpha_s\alpha_v}{M^4}\,\chi_s\left[ C_{P}\langle g^a,g^b|\,\tr
{G}_{\mu\nu}{\tilde{G}}^{\mu\nu}|\,0\rangle\,\langle 1^{+-}|\, P|
1^{--}\rangle  \ + \right.
\\ \left. + C_{L}\langle
g^a,g^b|\,\tr{G}_{\mu\nu}{G}_{\alpha\beta}|\,0\rangle  \langle 1^{+-}|\,
L^{\mu\nu\alpha\beta}| 1^{--}\rangle\right],
\end{multline*}
where the s-wave approximation implies the following form of the matrix elements;
\be \nonumber
\langle 1^{+-}|\, P|1^{--}\rangle = \frac{\epsilon^+\cdot\epsilon^-}{m_{1^{-}}}\, {\bf M_{1^{--}1^{+-}}^P},
\ee
\begin{multline*}
\langle 1^{+-},q|\,L_{\mu\nu\alpha\beta}| 1^{--},p\rangle = \frac{\bf M_{1^{--}1^{+-}}^L}{m_{1^-}^3}\,(\epsilon_{\mu\nu\rho\sigma}p^\rho {\epsilon^-}^\sigma(p_\alpha{\epsilon^+}_\beta - p_\beta{\epsilon^+}_\alpha ) +
\\ + \epsilon_{\alpha\beta\rho\sigma}p^\rho {\epsilon^-}^\sigma(p_\mu{\epsilon^+}_\nu - p_\nu{\epsilon^+}_\mu ) - {\rm traces} ) + \ldots \ ,
\end{multline*}
The above amplitude gives a decay rate
\begin{multline}
\label{1-to1+gg}
\Gamma_{1^{-}\to 1^{+}gg} =\frac{\alpha_s^2\alpha_v^2}{2^{9}\pi^3 M^8}\frac1{3}(N_c^2-1)\chi_s^2 m_{1^-}^{3}\left(4 C_L^2 ({\bf M_{1^{--}1^{+-}}^L})^2\,f_L(a) + \right.
\\ \left. + C_P^2({\bf M_{1^{--}1^{+-}}^P})^2\,f_P(a)\right)+\ldots \ ,
\end{multline}
where we define the dimensionless functions $f_L(a)$ and $f_P(a)$ of $a\equiv m_{1^{+}}^2/m_{1^-}^2$ as
\begin{multline*}
f_L(a)=-\frac{1}{15120 a} (171 a^7-1295 a^6+4410 a^5-9450 a^4+11025 a^3-
\\ - 4221 a^2-630 a-10) -\frac{1}{36}\, a (5 a-9) \log (a),
\end{multline*}
\begin{multline*}
f_P(a)= -\frac{1}{120 a} \left(a^6+36 a^5+1305 a^4-1305 a^2-36 a-1\right) +
\\ + \frac{1}{2}\, a \left(9 a^2+28 a+9\right) \log (a).
\end{multline*}
For the values of v-glueball masses from the spectrum in figure~\ref{fig spectrum}, $a\simeq 0.6$, $f_L(0.6)\simeq 3\times 10^{-5}$ and $f_P(0.6)\simeq 7\times 10^{-5}$.

The ratio of the decay rate~(\ref{1-to1+gg}) to the radiative two-body decay~(\ref{1--tog}) is
\be\nonumber
\frac{\Gamma_{1^{-}\to 1^{+}gg}}{\Gamma_{1^{-}\to 0^{+}\gamma}}=6\times 10^{-6}\frac{\alpha_s^2}{\alpha\alpha_v}\,\frac{\chi_s^2}{\chi^2}\,\left(\frac{4C_L^2({\bf M_{1^{--}1^{+-}}^L})^2+\frac{f_P}{f_L}C_P^2({\bf M_{1^{--}1^{+-}}^L})^2}{(\bf M_{1^{--}0^{++}}^{\Omega})^2}\right).
\ee

Unless there is an extreme degeneracy in the $SU(5)$ multiplet of $X$ particles,
which is unnatural due to $SU(5)$-asymmetric renormalization of the masses, we expect $\chi$ is at least of order $0.1$, so the coefficient in front of the ratio of the form-factors will be around $ 10^{-2}$ or smaller.

For other states in the $C$-odd sector, a rough estimate confirms that the ratio of the three-body decays to the radiative decays is never greater than $1/10$ and is typically much smaller. 
Thus, we conclude that the three-body decays in this sector are never dominant.  Since
most such decays are to gluons, and are therefore very difficult to observe, the three-body
processes can for current purposes be ignored.

\end{appendix}


\end{document}